\documentclass{ieeeaccess}
\usepackage{algorithmic}
\usepackage{amsmath,amssymb,amsfonts}
\usepackage{cite}
\usepackage{enumitem}
\usepackage{graphicx}
\usepackage{subcaption}
\usepackage{textcomp}
\usepackage{tikz}
\usepackage{tabularx}
\usepackage{booktabs}
\usetikzlibrary{shapes.geometric}
\usetikzlibrary{decorations.pathmorphing}

\NewSpotColorSpace{PANTONE}
\AddSpotColor{PANTONE} {PANTONE3015C} {PANTONE\SpotSpace 3015\SpotSpace C} {1 0.3 0 0.2}
\SetPageColorSpace{PANTONE}%

\def\BibTeX{{\rm B\kern-.05em{\sc i\kern-.025em b}\kern-.08em
    T\kern-.1667em\lower.7ex\hbox{E}\kern-.125emX}}
\begin{document}
\history{}
\doi{}

\title{Analytical model for the relation between signal bandwidth and spatial resolution in steered-response power phase transform (SRP-PHAT) maps}
\author{\uppercase{Guillermo Garc\'{\i}a-Barrios}\authorrefmark{1}, \uppercase{Juana M. Guti\'errez-Arriola}\authorrefmark{1}\IEEEmembership{Member, IEEE}, \uppercase{Nicol\'as S\'aenz-Lech\'on}\authorrefmark{1}, \uppercase{V\'{\i}ctor J. Osma-Ruiz}\authorrefmark{1}, \uppercase{Rub\'en Fraile}\authorrefmark{1}}
\address[1]{CITSEM, Universidad Polit\'ecnica de Madrid - Campus Sur. Alan Turing~3, 28031 Madrid. Spain (e-mail: guillermo.garcia.barrios@upm.es; juana.gutierrez.arriola@upm.es; nicolas.saenz@upm.es; v.osma@upm.es; r.fraile@upm.es)}
\tfootnote{This research was partially funded by the Universidad Polit\'ecnica de Madrid through its Programa Propio de I+D+I, specifically the Mentorship and Predoctoral calls. The authors gratefully acknowledge the Universidad Polit\'ecnica de Madrid for providing computing resources on Magerit Supercomputer.}

\markboth
{Garc\'{\i}a-Barrios \headeretal: Relation between signal bandwidth and spatial resolution in SRP-PHAT maps}
{Garc\'{\i}a-Barrios \headeretal: Relation between signal bandwidth and spatial resolution in SRP-PHAT maps}

\corresp{Corresponding author: Rub\'en Fraile (e-mail: r.fraile@upm.es).}

\begin{abstract}
An analysis of the relationship between the bandwidth of acoustic signals and the required resolution of steered-response power phase transform (SRP-PHAT) maps used for sound source localization is presented. This relationship does not rely on the far-field assumption, nor does it depend on any specific array topology. The proposed analysis considers the computation of a SRP map as a process of sampling a set of  generalized cross-correlation (GCC) functions, each one corresponding to a different microphone pair. From this approach, we derive a rule that relates GCC bandwidth with inter-microphone distance, resolution of the SRP map, and the potential position of the sound source relative to the array position. This rule is a sufficient condition for an aliasing-free calculation of the specified SRP-PHAT map. Simulation results show that limiting the bandwidth of the GCC according to such rule leads to significant reductions in sound source localization errors when sources are not in the immediate vicinity of the microphone array. These error reductions are more relevant for coarser resolutions of the SRP map, and they happen in both anechoic and reverberant environments.
\end{abstract}


\titlepgskip=-15pt

\maketitle

\section{Introduction}
\PARstart{S}{ound} source localization based on steered-response power (SRP) maps computed using the generalized cross-correlation (GCC) function with phase transform (PHAT), i.e. SRP-PHAT, has been reported to perform robustly against noise and, especially, reverberation \cite{CCTM16,DiSB01}. The PHAT applied to the GCC function has the effect of narrowing its maxima, hence allowing a more precise identification of the time difference of arrival (TDOA) between microphones \cite{KnCa76}. However, this increased precision can only be exploited by correspondingly reducing the spatial resolution\footnote{Herein, resolution is defined as the distance between contiguous points in the map grid. Therefore, the lowest resolutions correspond to the finest map grids, and the highest resolutions are associated with the coarsest grids.} of SRP maps, which turns out to be one of the main drawbacks of sound source localization based on SRP-PHAT \cite{CCTM16} since it implies higher computational costs. 

Therefore, implementing a sound source localization system based on SRP involves finding a balance between computational cost and precision. To present, this challenge has been approached in several ways. One of them has consisted in performing calculations at several resolution levels, from coarsest to finest, and limiting the extent of the map each time the resolution is decreased. This hierarchical search can be implemented, for instance, by defining rectangular and regular grids whose cells are iteratively decomposed into finer grids \cite{ZoDu04, MCLE13, NMLB14, LMNB15}. Instead of conducting the hierarchical search using  regular grids, some researchers have proposed grouping regions by TDOA \cite{YoLC16}, or decreasing resolution mainly in regions where the SRP function is expected to vary most abruptly \cite{SaDF18}. Other approaches try to avoid iterative processes, thus keeping resolution fixed, while computational cost is maintained at an affordable level by restricting the search space only to regions where the sound source is expected to be, according to some \emph{a priori} information \cite{AsBP15, AHTS20}. 

When coarse spatial resolutions are used for generating  SRP maps based on spiky functions such as the GCC-PHAT, two risks are taken. On the one hand, the narrow peak corresponding to the global maximum of the GCC may not be adequately sampled; on the other hand, spurious local maxima of the GCC may be reflected in the SRP maps. These two effects can distort localization estimates, an effect that is more likely to happen at the first stages of hierarchical searches, thus leading to severe errors in the overall results. In order to avoid such errors several approaches have been proposed so far, such as stochastic region contraction (SRC), which involves performing a stochastic search of the highest peaks in the SRP map \cite{DoSY07} before decreasing map resolution and reducing map extent; calculating the integral of the GCC-PHAT along an interval of time delay values defined by the position of each grid point and the spatial resolution of the map \cite{CoML11}; or designing the map grid considering the specific geometry of the microphone array \cite{SaDF17}. Alternative approaches based on deep learning have also been proposed to reduce the number of local maxima in the SRP map by either post-processing the GCC \cite{VePM21}, or the map itself \cite{DiMB21}.

Qualitatively, the width of peaks in the GCC are known to be related to the spectral content of the audio signal. Thus, signal spectrum or, more specifically, signal bandwidth is not independent of the spatial map resolution required to obtain good localization estimates. This relation can be used to design the afore-mentioned hierarchical search considering the bandwidth of the specific signal being processed \cite{ZoDu04}, and it is also implicit in proposals such as integrating the GCC-PHAT \cite{CoML11} (integration is equivalent to low-pass filtering) or applying multi-band analysis to reduce the effects of spatial aliasing \cite{FIAD19}.

The peak narrowing in the GCC becomes particularly relevant for large inter-microphone distances when the sound source is likely to be near the microphone array, or even inside it. For this reason, distributed microphone arrays potentially allow better precision in source localization \cite{SiPa99}, but at the cost of higher computational load, as reasoned before. For these particular cases, a quantitative rule relating signal bandwidth and inter-microphone distances has been proposed in order to avoid the appearance of spurious secondary lobes in the beam pattern of the array. Specifically, it is commonly assumed that the acoustic wavelength for far-field measurements using microphone arrays should be larger than twice the inter-microphone distance  (e.g. \cite{McKu12}). However, such rule does not consider map resolution.

Considering all the previous questions as a whole, it is straightforward to conclude that there is a relationship between inter-microphone distance or array size, signal bandwidth, and the spatial resolution required to avoid under-sampling the GCC. In this paper, we present a rule that quantifies this relation. The analysis leading to this rule does not rely on the far-field assumption and it is not dependent on any specific array topology. The rule can be applied to hierarchical searches at every resolution level to avoid the emergence of spurious maxima at the corresponding SRP maps, hence achieving lower errors in sound source localization estimates. Furthermore, it provides an alternative interpretation, based on basic signal processing theory, of algorithms involving GCC integration \cite{CoML11}, design of map grids with reduced resolution in certain areas \cite{SaDF17}, or adjustment of grid resolution as a function of signal bandwidth \cite{ZoDu04}.

The adopted approach considers the computation of a SRP map as a process of sampling a set of GCC functions, each one corresponding to a different microphone pair. This theoretical analysis is presented in sections \ref{sec:ProblemStatement} and \ref{sec:Geometrical analysis}, while its implications for SRP map calculations are discussed in section \ref{sec:Implications}. Section \ref{sec:Calculation} shows how to incorporate the previous theoretical results into the process of calculating SRP maps by limiting the bandwidth of the GCC specifically for each point in the map. Results obtained using this approach presented in section \ref{sec:Experiments} indicate that it can provide significant error reductions in the estimation of source positions. The conditions in which such improvements can be achieved are discussed in section \ref{sec:Discussion}.

\section{Problem statement}
\label{sec:ProblemStatement}
Sound source localization consists in estimating the position $\vec{r}_\mathrm{s}$ of an acoustic source with respect to a certain coordinate reference, given the corresponding acoustic signals captured at a set of $K$ microphones whose positions are known. When a SRP algorithm is used, the source position is estimated as \cite{DiSB01}:
\begin{equation}
	\vec{r}_\mathrm{s} \approx \arg\max P\left(\vec{r}\right),
\end{equation} 
where $P\left(\vec{r}\right)$ is the value of the SRP map at position $\vec{r}$. This can be calculated as:
\begin{IEEEeqnarray}{r,c,l}
	P\left(\vec{r}\right) &=& 2\pi \sum_{k=1}^K \sum_{l=1}^K R_{kl}\left( \tau_l\left(\vec{r}\right) - \tau_k\left(\vec{r}\right) \right)  \label{eq:SRP}\\
	&=&2\pi \sum_{k=1}^K \sum_{l=1}^K R_{kl}\left( \tau_{kl}\left(\vec{r}\right)\right),\nonumber
\end{IEEEeqnarray} 
where $\tau_k\left(\vec{r}\right)$ is the propagation delay between position $\vec{r}$ and the position of the $k^\mathrm{th}$ microphone, and $R_{kl}\left(\tau_{kl}\left(\vec{r}\right)\right)$ is the GCC function between the sound signals captured at microphones $k$ and $l$, respectively $s_k\left(t\right)$ and $s_l\left(t\right)$, evaluated at time lag $\tau_{kl}\left(\vec{r}\right)$. When the PHAT weighting is used, the GCC can be calculated as:
\begin{equation}
	R_{kl}\left(\tau\right) = \int_{-\infty}^{\infty} \frac{S_k\left(\omega\right)S_l^*\left(\omega\right)\cdot \mathrm{e}^{j\omega \tau}}{2\pi\left|S_k\left(\omega\right)S_l^*\left(\omega\right)\right|} \mathrm{d}\omega,
	\label{eq:GCCPHAT}
\end{equation}
being $S_k\left(\omega\right)$ the Fourier transform of $s_k\left(t\right)$, and $j$ the imaginary unit. This calculation is problematic when the integral spans over frequencies for which the signal-to-noise ratio (SNR) corresponding to $s_k\left(t\right)$ and $s_l\left(t\right)$ is low \cite{KnCa76}, due to the division in (\ref{eq:GCCPHAT}). In the case of passband signals, whose SNR is high only within a certain frequency interval $\omega_\mathrm{min} \le \omega \le \omega_\mathrm{max}$, this can be solved by limiting the integration to the same interval:
\begin{equation}
	R_{kl}\left(\tau\right) = \int_{\omega_\mathrm{min} \le \left|\omega \right|\le \omega_\mathrm{max}} \frac{S_k\left(\omega\right)S_l^*\left(\omega\right)\cdot \mathrm{e}^{j\omega \tau}}{2\pi\left|S_k\left(\omega\right)S_l^*\left(\omega\right)\right|} \mathrm{d}\omega.
	\label{eq:BandlimitedGCCPHAT}
\end{equation}

Therefore, $P\left(\vec{r}\right)$ is a non-linear function of a three-dimensional variable $\vec{r}$. Its maximization is commonly performed by evaluating it on a set of predefined points (usually a grid) in the area of interest, and selecting the point yielding the highest value \cite{DiSB01}. Considering (\ref{eq:SRP}), this approach can be seen as a sampling of $P\left(\vec{r}\right)$ in which each sample is obtained by combining certain samples of the GCC functions $R_{kl}\left(\tau\right)$. When jumping from one of these predefined points $\vec{r}$ to a contiguous one in the grid $\vec{r}+\Delta \vec{r}$, the time lags at which the GCC functions need to be evaluated change from $\tau_{kl}\left(\vec{r}\right)$ to $\tau_{kl}\left(\vec{r}+\Delta \vec{r}\right)$, hence missing all intermediate values of the GCC functions. In cases for which $\left|\tau_{kl}\left(\vec{r}\right) - \tau_{kl}\left(\vec{r}+\Delta \vec{r}\right)\right|$ is large enough, some narrow peaks of the GCC may be missed, leading to localization errors like those illustrated in \cite[Figs. 7 and 8]{VMMP16}.

According to this approach, the calculation of SRP maps can be understood as a compound sampling process of the GCC functions corresponding to all microphone pairs. The research
question faced here is whether some basic sampling theory can be applied to model this process and derive an equation that relates GCC bandwith to grid resolution, and whether such a model could be useful for improving the localization performance of SRP-PHAT algorithms by modifying the calculation of GCC functions instead of making use of GCC integration at a later stage as in \cite{CoML11}, or designing point grids specific for each scenario, like in \cite{SaDF17}.

\section{Geometrical analysis}
\label{sec:Geometrical analysis}
Let's consider the simple case of two microphones $k$ and $l$ and one point $\vec{r}= \left(x, y, z\right)$ for which $P\left(\vec{r}\right)$ needs to be evaluated (Fig.~\ref{fig:Framework}). Without loosing generality, let's further suppose that both microphones are symmetrically arranged around the origin of coordinates, so microphone $k$ is placed at position $\vec{r}_\mathrm{m} = \left(x_\mathrm{m}, y_\mathrm{m}, z_\mathrm{m}\right)$, and the position of microphone $l$ is $-\vec{r}_\mathrm{m}$. Given the value of the sound velocity $c$, the TDOA between microphones $k$ and $l$ associated to point $\vec{r}$ is:

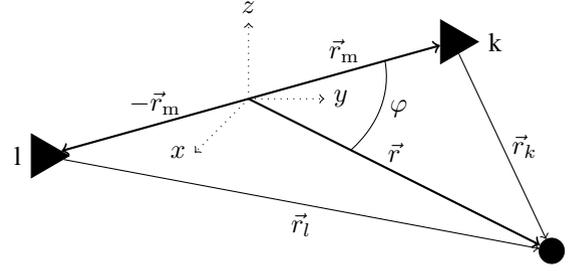
\begin{figure}[t]
	\centering
	\begin{tikzpicture}[
		triangle/.style = {regular polygon, regular polygon sides=3, shape border rotate=30},
		receiver/.style={draw=black,fill=black}]
		\node (source) at (4,-2) [circle,draw=black,fill=black]{};
		\node [triangle, receiver, label=right:k] (micro1) at (2.7,0.75){};
		\node [triangle, receiver,label=left:l] (micro2) at (-2.7,-0.75){};
		
		\draw[->, dotted] (0,0) -- (1,0) node [auto, right] {$y$};
		\draw[->, dotted] (0,0) -- (-0.707,-0.707) node [auto, left] {$x$};
		\draw[->, dotted] (0,0) -- (0,1) node [auto, above] {$z$};
		\draw[->, thick] (0,0) -- (micro1) node [above,midway] {$\vec{r}_\mathrm{m}$};
		\draw[->, thick] (0,0) -- (micro2) node [above,midway] {$-\vec{r}_\mathrm{m}$};
		\draw[->, thick] (0,0) -- (source) node [above,midway] {$\vec{r}$};
		\draw (1.8,0.5) arc(8:-40:2 and 1.5) node [right,midway] {$\varphi$};
		\draw[->,thin] (micro1) -- (source) node [right,midway] {$\vec{r}_k$};
		\draw[->,thin] (micro2) -- (source) node [below,midway] {$\vec{r}_l$};
	\end{tikzpicture}
	\caption{Simplified scenario comprising two microphones (triangles) and one position (circle).}
	\label{fig:Framework}
\end{figure}

\begin{equation}
	\tau_{kl}\left(\vec{r}\right) = \frac{1}{c}\left(r_l -r_k\right) =  \frac{1}{c}\left(\left\| \vec{r}+\vec{r}_\mathrm{m}\right\| - \left\| \vec{r}-\vec{r}_\mathrm{m}\right\| \right),
	\label{eq:TDOA}
\end{equation}
where $\left\| \cdot \right\|$ is the Euclidean norm, and $r_k = \left\| \vec{r}_k \right\|$. We are interested in studying the sampling process of 
$R_{kl}\left(\tau\right)$, so we analyze how the sampling time $\tau_{kl}$ changes as the potential source position changes from one grid point to a contiguous one. Specifically, according to a first-order Taylor approximation, given $\tau_{kl}\left(\vec{r}\right)$, the TDOA at a contiguous point $\vec{r} + \Delta \vec{r}$ can be approximated as \cite[chap.11]{WrSp10}:
\begin{equation}
	\tau_{kl}\left(\vec{r} + \Delta \vec{r}\right) \approx \tau_{kl}\left(\vec{r}\right) + \nabla \tau_{kl}\left(\vec{r}\right) \cdot \Delta \vec{r},
\end{equation}
where $\nabla \tau_{kl}\left(\vec{r}\right)$ is the gradient of the TDOA and $\cdot$ is the dot product. The interval between adjacent samples of $R_{kl}\left(\tau\right)$ can then be estimated as:
\begin{equation}
	\Delta \tau_{kl} = \left|\tau_{kl}\left(\vec{r} + \Delta \vec{r}\right) - \tau_{kl}\left(\vec{r}\right)\right| \approx \left| \nabla \tau_{kl}\left(\vec{r}\right) \cdot \Delta \vec{r} \right|.
\end{equation}
According to the properties of the dot product:
\begin{equation}
	\Delta \tau_{kl} \approx \left| \nabla \tau_{kl}\left(\vec{r}\right) \cdot \Delta \vec{r} \right| \le \left\|\nabla \tau_{kl}\left(\vec{r}\right)\right\| \cdot \Delta r,
	\label{eq:SamplingTime}
\end{equation}
where $\Delta r = \left\| \Delta \vec{r} \right\|$. Therefore, the maximum sampling interval of $R_{kl}\left(\tau\right)$ is bounded by the product between the resolution of the SRP map and the modulus of the gradient of the TDOA. The resolution of the SRP map is defined as the distance between any point in the map grid and its closest surrounding points. It is mathematically represented by $\Delta r$, previously defined as the distance between contiguous points in the grid. This resolution is constant for regular grids, and position-dependent for irregular grids. In what follows, no assumption is made with respect to this issue. That is, the subsequent formulation is valid for both regular and irregular grids.   

Considering (\ref{eq:TDOA}), the gradient of the TDOA can be calculated as:
\begin{equation}
	\nabla \tau_{kl}\left(\vec{r}\right) = \left(\frac{\partial \tau_{kl}}{\partial x},  \frac{\partial \tau_{kl}}{\partial y}, \frac{\partial \tau_{kl}}{\partial z}\right),
\end{equation}
with
\begin{IEEEeqnarray}{r,c,l}
	\frac{\partial \tau_{kl}}{\partial x} &=& \frac{1}{c}\left(\frac{x+x_\mathrm{m}}{\left\| \vec{r}+\vec{r}_\mathrm{m}\right\|}-\frac{x-x_\mathrm{m}}{\left\| \vec{r}-\vec{r}_\mathrm{m}\right\|}\right), \nonumber \\
	\frac{\partial \tau_{kl}}{\partial y} &=& \frac{1}{c}\left(\frac{y+y_\mathrm{m}}{\left\| \vec{r}+\vec{r}_\mathrm{m}\right\|}-\frac{y-y_\mathrm{m}}{\left\| \vec{r}-\vec{r}_\mathrm{m}\right\|}\right),\nonumber \\
	\frac{\partial \tau_{kl}}{\partial z} &=& \frac{1}{c}\left(\frac{z+z_\mathrm{m}}{\left\| \vec{r}+\vec{r}_\mathrm{m}\right\|}-\frac{z-z_\mathrm{m}}{\left\| \vec{r}-\vec{r}_\mathrm{m}\right\|}\right).\nonumber 
\end{IEEEeqnarray}
And the square of its Euclidean norm is: 
\begin{IEEEeqnarray}{r,c,l}
	\left\|\nabla \tau_{kl}\left(\vec{r}\right)\right\|^2&=&
	\left(\frac{\partial \tau_{kl}}{\partial x}\right) ^2 +\left(\frac{\partial \tau_{kl}}{\partial y}\right)^2 + \left(\frac{\partial \tau_{kl}}{\partial z}\right)^2 = \nonumber\\
	&=& \frac{1}{c^2}\left(\frac{\left\| \vec{r}+\vec{r}_\mathrm{m}\right\|^2}{\left\| \vec{r}+\vec{r}_\mathrm{m}\right\|^2} + \frac{\left\| \vec{r}-\vec{r}_\mathrm{m}\right\|^2}{\left\| \vec{r}-\vec{r}_\mathrm{m}\right\|^2}-\right. \label{eq:GradientSquareNorm}\\
	&&\left.-2 \frac{x^2-x_\mathrm{m}^2+y^2-y_\mathrm{m}^2+z^2-z_\mathrm{m}^2}{\left\| \vec{r}+\vec{r}_\mathrm{m}\right\|\cdot \left\| \vec{r}-\vec{r}_\mathrm{m}\right\|}\right) =  \nonumber \\
	&=&\frac{2}{c^2}\left(1- \frac{r^2-r_\mathrm{m}^2}{\left\| \vec{r}+\vec{r}_\mathrm{m}\right\|\cdot \left\| \vec{r}-\vec{r}_\mathrm{m}\right\|}\right),\nonumber	
\end{IEEEeqnarray}
where, similarly as before, $r = \left\| \vec{r} \right\|$ and $r_\mathrm{m} = \left\| \vec{r}_\mathrm{m} \right\|$. According to the law of cosines:
\begin{equation}
	r_k = \left\|\vec{r}-\vec{r}_\mathrm{m}\right\| =\sqrt{r^2 + r_\mathrm{m}^2 -2 r r_\mathrm{m} \cos\varphi},
	\label{eq:CosineLaw1}
\end{equation}
where $\varphi$ is the angle indicated in Fig.~\ref{fig:Framework}. Analogously:
\begin{equation}
	r_l = \left\|\vec{r}+\vec{r}_\mathrm{m}\right\| =\sqrt{r^2 + r_\mathrm{m}^2 +2 r r_\mathrm{m} \cos\varphi}.
	\label{eq:CosineLaw2}
\end{equation}
Now, substituting (\ref{eq:CosineLaw1}) and (\ref{eq:CosineLaw2}) in (\ref{eq:GradientSquareNorm}):
\begin{IEEEeqnarray}{r,c,l}
	\left\|\nabla \tau_{kl}\left(\vec{r}\right)\right\|^2&=& \frac{2}{c^2}\left(1- \frac{r^2-r_\mathrm{m}^2}{\sqrt{r^2 + r_\mathrm{m}^2 +2 r r_\mathrm{m} \cos\varphi}} \cdot \right. \nonumber \\
	&& \left. \cdot \frac{1}{\sqrt{r^2 + r_\mathrm{m}^2 -2 r r_\mathrm{m} \cos\varphi}}\right) =  \\
	&=& \frac{2}{c^2}\left(1- \frac{r^2-r_\mathrm{m}^2}{\sqrt{\left(r^2 + r_\mathrm{m}^2\right)^2 -4 r^2 r_\mathrm{m}^2 \cos^2\varphi}}\right).\nonumber
\end{IEEEeqnarray}
Thus, the Euclidean norm of the gradient is:
\begin{IEEEeqnarray}{r,c,l}
	\left\|\nabla \tau_{kl}\left(\vec{r}\right)\right\| &=& \frac{1}{c}\sqrt{2- \frac{2\left(r^2-r_\mathrm{m}^2\right)}{\sqrt{\left(r^2 + r_\mathrm{m}^2\right)^2 -4 r^2 r_\mathrm{m}^2 \cos^2\varphi}}} = \label{eq:GradientNorm}\\
	&=&\frac{1}{c}\sqrt{2- \frac{2\left(\left(\frac{r}{r_\mathrm{m}}\right)^2-1\right)}{\sqrt{\left(\left(\frac{r}{r_\mathrm{m}}\right)^2 + 1\right)^2 -4 \left(\frac{r}{r_\mathrm{m}}\right)^2 \cos^2\varphi}}}.\nonumber
\end{IEEEeqnarray}

\begin{figure}[t!]
	\centering
	\includegraphics[width=\linewidth]{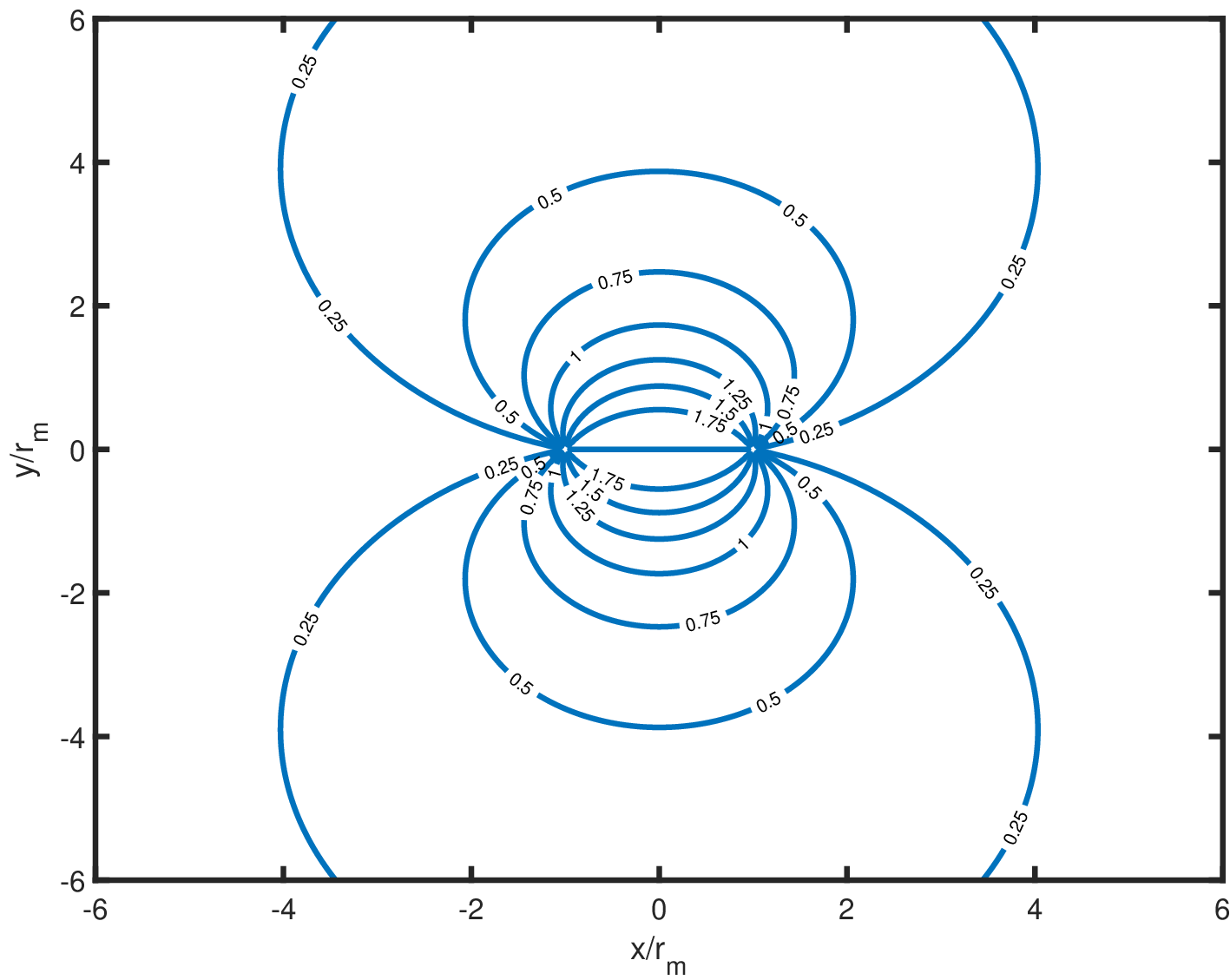}
	\caption{Contour plot of $c\left\|\nabla \tau_{kl}\left(\vec{r}\right)\right\|$ in the horizontal plane when both microphones are in that plane.}
	\label{fig:GradientMap}
\end{figure}

\begin{figure}[t!]
	\centering
	\includegraphics[width=\linewidth]{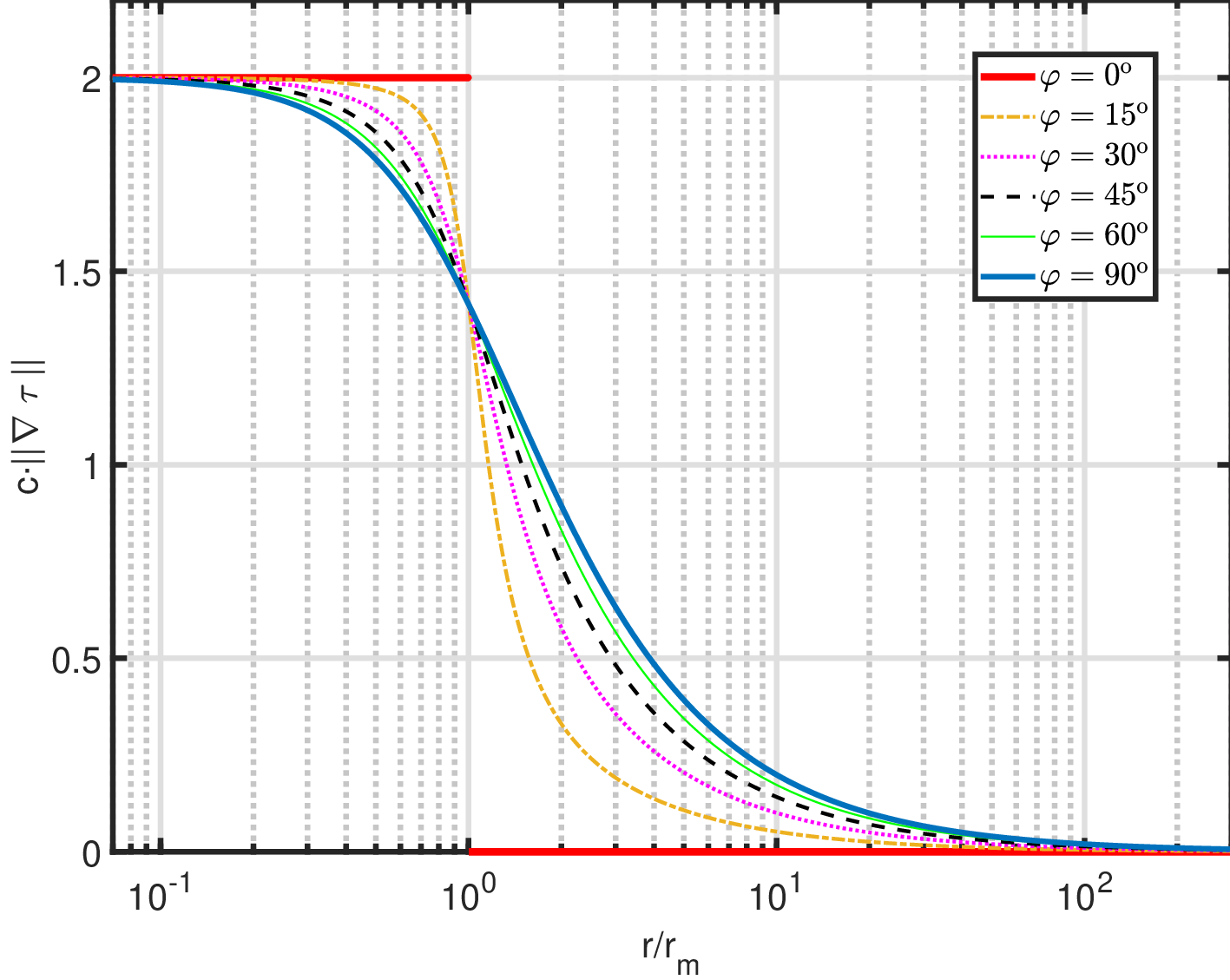}
	\caption{Plot of $c\left\|\nabla \tau_{kl}\left(\vec{r}\right)\right\|$ as a function of distance for several angles.}
	\label{fig:Gradient}
\end{figure}

This expression shows that the norm of the gradient depends on the distance to the centre of the microphone array, relative to half the inter-microphone distance $\left(\frac{r}{r_\mathrm{m}}\right)$, and on the angle $\varphi$. The contour plot in Fig.~\ref{fig:GradientMap} shows that the largest gradients occur near the centre of the array and for angles near $90^\mathrm{o}$, with a maximum at the segment linking both microphones. This result is consistent with the simulation results on SRP sensitivity illustrated in \cite{SaDF17}. Fig.~\ref{fig:Gradient} depicts the relation between the norm of the gradient and $\left(\frac{r}{r_\mathrm{m}}\right)$ for several angles. This graph  shows that the maximum value of $c\left\|\nabla \tau_{kl}\left(\vec{r}\right)\right\|$ is 2, which happens between both microphones, and that the largest differences for diverse values of $\varphi$ happen approximately for $0.2<\left(\frac{r}{r_\mathrm{m}}\right)<20$, i.e. for cases in which the difference between the size of the array and the distance between the array itself and the source positions is one order of magnitude at most. On the opposite, when $\left(\frac{r}{r_\mathrm{m}}\right)$ is large (far field) the influence of $\varphi$ vanishes.

\section{Implications for the calculation of the SRP map}
\label{sec:Implications}
According to the approach introduced in the previous section, the calculation of a SRP-PHAT map (\ref{eq:SRP}) basically consists in a sample-and-sum process that includes sampling of several GCC-PHAT functions (\ref{eq:BandlimitedGCCPHAT}) with variable sampling intervals (\ref{eq:SamplingTime}) whose values depend on the resolution of the SRP map, on the specific position being evaluated, and on the microphone positions. This sample-and-sum process leads to erroneous results when the selected samples of the GCC cannot represent some narrow peaks of the function. As stated by the sampling theorem \cite[chap.8]{OpWY83}, if such loss of information (due to aliasing) is to be avoided, then the inverse of the sampling time should be greater than twice the bandwidth of the signal:
\begin{equation}
	2\cdot \frac{\omega_\mathrm{max}}{2\pi} < \frac{1}{\Delta \tau_{kl}} \ \Longrightarrow \ \Delta \tau_{kl} < \frac{\pi}{\omega_\mathrm{max}}, 
	\label{eq:SamplingCondition1}
\end{equation}
where it has been implicitly assumed that bandpass sampling is not feasible or, in other words, that $\omega_\mathrm{max} > 2\cdot \omega_\mathrm{min}$. If condition (\ref{eq:SamplingCondition1}) is to be met in all cases, then taking (\ref{eq:SamplingTime}) into account one can derive a sufficient condition that allows obtaining a SRP map that does not suffer from aliasing in the sampling of GCCs, given a specific microphone array and the corresponding audio signals:
\begin{equation}
	\left\|\nabla \tau_{kl}\left(\vec{r}\right)\right\| \cdot \Delta r < \frac{\pi}{\omega_\mathrm{max}}. 
	\label{eq:SamplingCondition2}
\end{equation}

This relationship between distance $r$, array size $r_\mathrm{m}$ (both implicit in $\left\|\nabla \tau_{kl}\left(\vec{r}\right)\right\|$), map resolution $\Delta r$, and signal bandwidth $\omega_\mathrm{max}$ can be exploited in several ways, depending on
which of these magnitudes are defined by the scenario where the localization system is to be deployed and which ones are adjustable:
\begin{itemize}
	\item For distributed microphone arrays in which the sound source is likely to be placed somewhere between the microphones, this implies $r\lesssim r_\mathrm{m}$ and in this case $\sqrt{2} \lesssim c\left\|\nabla \tau_{kl}\left(\vec{r}\right)\right\| \le 2$ (see Fig.~\ref{fig:Gradient}). Therefore, the required map resolution is:
	\begin{equation}
		\Delta r < \frac{c\pi}{2\cdot \omega_\mathrm{max}}. 
	\end{equation}
	\item When the distance from the source to the array is known to be larger than the array size, then the TDOA gradient is bounded by the case $\varphi = 90^\mathrm{o}$ (see Fig.~\ref{fig:Gradient}), thus: 
	\begin{equation}
		\Delta r \sqrt{1- \frac{\left(\frac{r}{r_\mathrm{m}}\right)^2-1}{\left(\frac{r}{r_\mathrm{m}}\right)^2 + 1}}< \frac{c\pi}{\sqrt{2}\cdot \omega_\mathrm{max}}, 
	\end{equation}
	and the map resolution can be estimated from the minimum value expected for $r$ or, alternatively, a map resolution dependent on $r$ can be set.
	\item In any of both cases, if the map resolution is not adjustable, the corresponding conditions can be used for setting the upper limit in the integral used for calculating the GCC (\ref{eq:BandlimitedGCCPHAT}).
	\item Alternatively, condition (\ref{eq:SamplingCondition2}) can also be used to calculate a SRP map with predefined resolution and variable GCC bandwidth. This implies limiting the GCC bandwidth in those points of the SRP map for which the TDOA gradient is high. Note that the time-domain effect of reducing the GCC bandwidth is similar to that of an integration of the GCC function, which is the operation proposed in \cite{CoML11}. Further details about this approach are given in the next section.
\end{itemize}

\section{Calculation of SRP maps with variable GCC bandwidth}
\label{sec:Calculation}

As pointed out before, when SRP maps with predefined resolution are to be generated, condition (\ref{eq:SamplingCondition2}) can be used to generate them while avoiding aliasing in the sampling of GCC functions. This can be done following the next procedure:
\begin{enumerate}
	\item Obtain the coordinates of the points in the grid used for generating the SRP map. Such grid will typically be characterized by its boundaries and a certain resolution $\Delta r$.
	\item For each point in the grid, the SRP will be obtained as the summation in (\ref{eq:SRP}). After initializing this summation, the next actions should be performed for each pair of microphones in the array $\left(k,l\right)$:
	\begin{enumerate}
		\item Obtain the vectors linking the grid point and the microphone positions $\vec{r}_k$ and $\vec{r}_l$ (see Fig.~\ref{fig:Framework}).
		\item Calculate $\vec{r}=0.5\cdot \left( \vec{r}_l + \vec{r}_k\right)$ and $\vec{r}_\mathrm{m}=0.5\cdot \left(\vec{r}_l - \vec{r}_k\right)$.
		\item Also compute $\cos\varphi = \left(\vec{r} \cdot \vec{r}_\mathrm{m} \right)/ \left(\left\|\vec{r}\right\| \cdot \left\|\vec{r}_\mathrm{m}\right\|\right)$.
		\item Use the previous results to calculate the norm of the gradient of the TDOA, as in (\ref{eq:GradientNorm}).
		\item Knowing $\Delta r$ and $\left\|\nabla \tau_{kl}\left(\vec{r}\right)\right\|$, estimate the maximum frequency $\hat{\omega}_\mathrm{max}$ that guarantees avoiding aliasing according to (\ref{eq:SamplingCondition2}).
		\item Calculate the GCC-PHAT as in (\ref{eq:BandlimitedGCCPHAT}), setting the upper limit to the integral equal to $\hat{\omega}_\mathrm{max}$.
	\end{enumerate}		
	\begin{equation}
		\hat{R}_{kl}\left(\tau\right) = \int_{\omega_\mathrm{min} \le \left|\omega \right|\le \hat{\omega}_\mathrm{max}} \frac{S_k\left(\omega\right)S_l^*\left(\omega\right)\cdot \mathrm{e}^{j\omega \tau}}{2\pi\left|S_k\left(\omega\right)S_l^*\left(\omega\right)\right|} \mathrm{d}\omega.
		\label{eq:BandlimitedGCCPHAT_2}
	\end{equation}
	\begin{enumerate}[resume]		
		\item Evaluate the resulting GCC for $\tau_{kl} = c\left(\left\|\vec{r}_l\right\| - \left\|\vec{r}_k\right\|\right)$, and add the result to the SRP value.
	\end{enumerate}
\end{enumerate}
\begin{figure}[t]
	\centering
	\includegraphics[width=\linewidth]{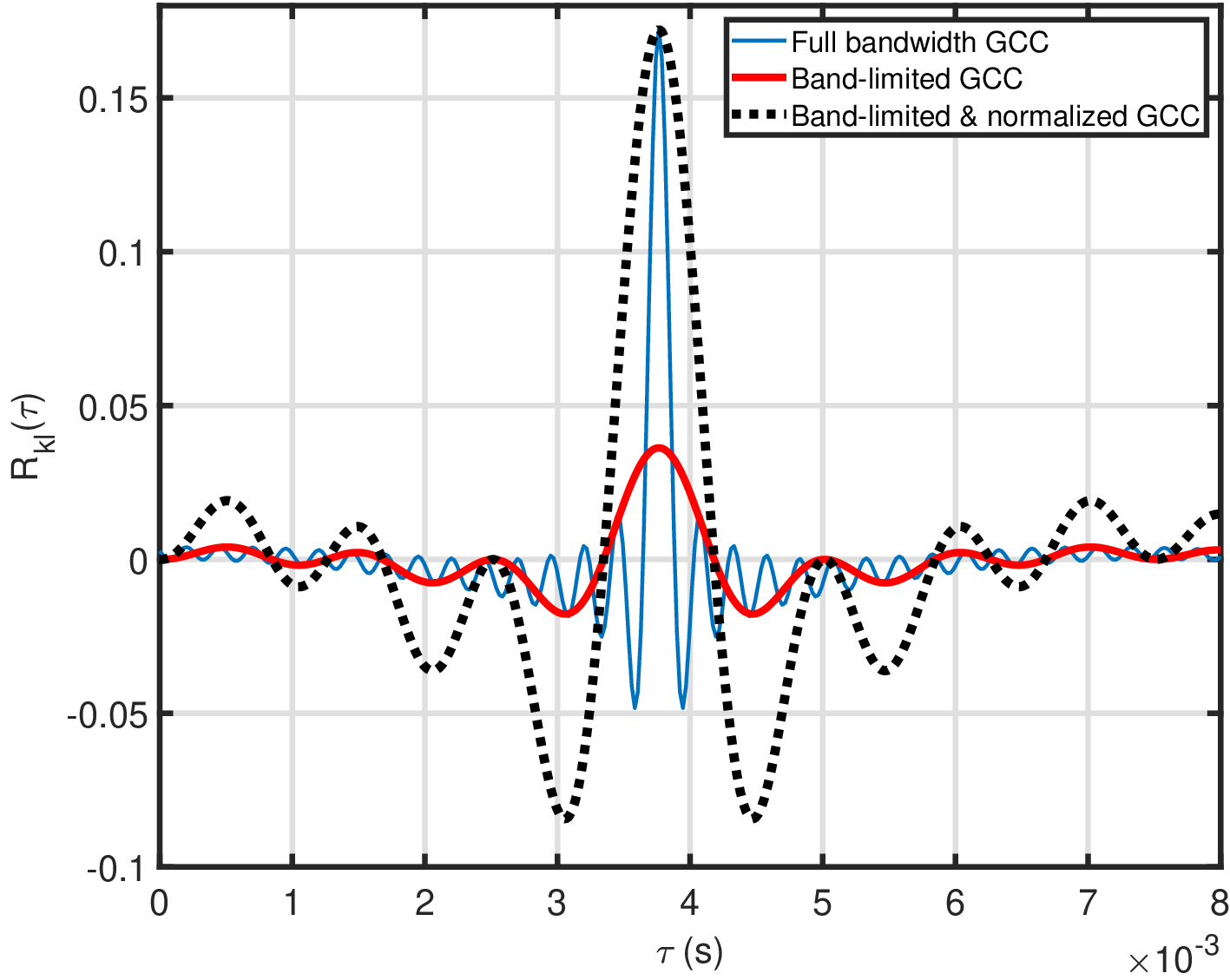}
	\caption{Cross correlation (GCC-PHAT) between two exactly equal speech signals taken from the dataset described in section \ref{subsec:AudioData}, having a 3.76~ms delay between them. The thin line depicts the GCC-PHAT calculated by integration along the 200~Hz-4000~Hz band, the continuous thick line shows the result of limiting this interval to 200~Hz-1000~Hz, and the dotted line shows the effect of applying the proposed normalization to the band-limited GCC-PHAT.}
	\label{fig:GCC}
\end{figure}

The SRP-PHAT function in (\ref{eq:SRP}) can be interpreted as a likelihood function that should be maximized to find the best possible estimate for the sound source position \cite{ZFBZ08}. Fig.~\ref{fig:GCC} illustrates the effect that the bandwidth limitation specified in step 2.f has on the resulting GCC (band-limited GCC). Apart from the expected effect of reducing the frequency of the oscillations in the GCC, and increasing the width of its main peak, limiting the bandwidth has the consequence of reducing the height of that peak. While increasing the width of the peak has the positive effect of reducing the aliasing when the GCC is sampled to generate a SRP map, reducing its height may reduce the likelihood associated to the true source position when evaluating (\ref{eq:SRP}), thus altering the position of the maximum value of the SRP map. This reduction on the amplitude of the GCC peak can be compensated by normalizing the band-limited GCC proportionally to the bandwidth reduction, as follows :
\begin{IEEEeqnarray}{r,c,l}
	\tilde{R}_{kl}\left(\tau\right) &=& \frac{\omega_\mathrm{max} - \omega_\mathrm{min}}{\hat{\omega}_\mathrm{max} - \omega_\mathrm{min}} \cdot  \label{eq:BandlimitedNormalisedGCCPHAT}\\
		&& \cdot \int_{\omega_\mathrm{min} \le \left|\omega \right|\le \hat{\omega}_\mathrm{max}} \frac{S_k\left(\omega\right)S_l^*\left(\omega\right)\cdot \mathrm{e}^{j\omega \tau}}{2\pi\left|S_k\left(\omega\right)S_l^*\left(\omega\right)\right|} \mathrm{d}\omega \nonumber,
\end{IEEEeqnarray}
where $\omega_\mathrm{min}$ and $\omega_\mathrm{max}$ are the limits of the signal bandwidth, and $\hat{\omega}_\mathrm{max}$ is the maximum frequency estimated in step 2.e. This normalization has the effect of keeping the value of the GCC peak unaltered, as shown in Fig.~\ref{fig:GCC}, at the cost of amplifying the oscillations of the function when $\tau$ moves away from the peak position.

\begin{figure*}[t]
	\centering
	\includegraphics[width=\linewidth]{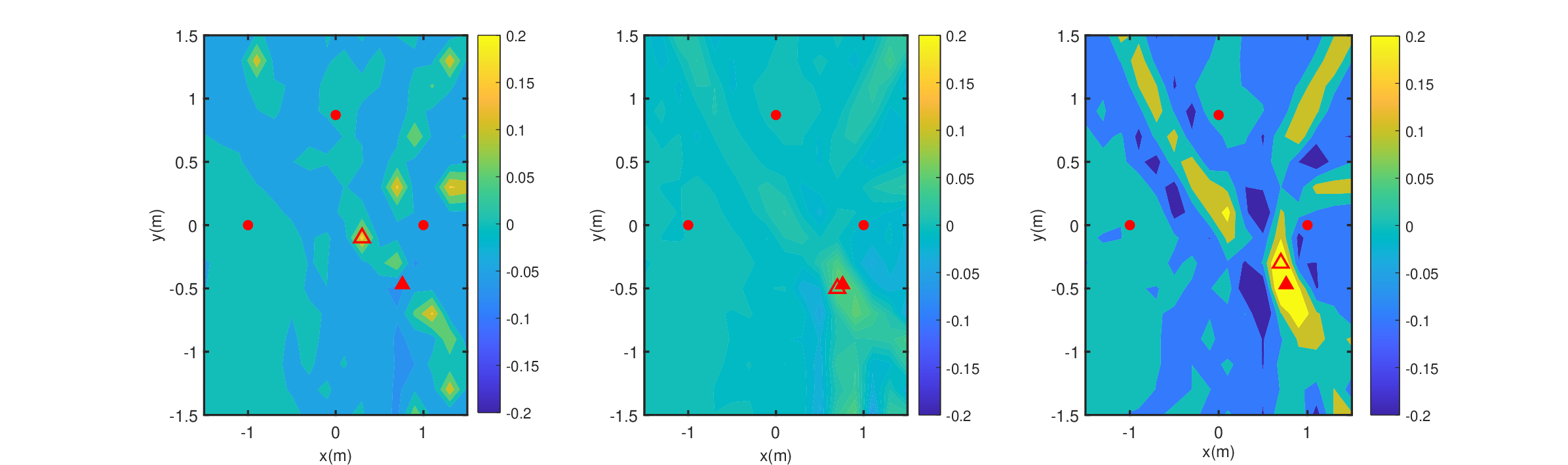}
	\caption{SRP-PHAT maps generated according to the standard procedure (left), applying (\ref{eq:BandlimitedGCCPHAT_2}) for limiting the bandwidth of the GCC (middle), and adding the normalization in (\ref{eq:BandlimitedNormalisedGCCPHAT}) (right). Red points indicate the simulated microphone positions, the filled triangles mark the simulated source position, and the empty triangles show the maximum peaks of the SRP maps, i.e. the estimated source positions. Anechoic conditions have been assumed, and the audio signal used for simulation is the same as in Fig.~ \ref{fig:GCC}.}
	\label{fig:SRP_near}
\end{figure*}

\begin{figure*}[t]
	\centering
	\includegraphics[width=\linewidth]{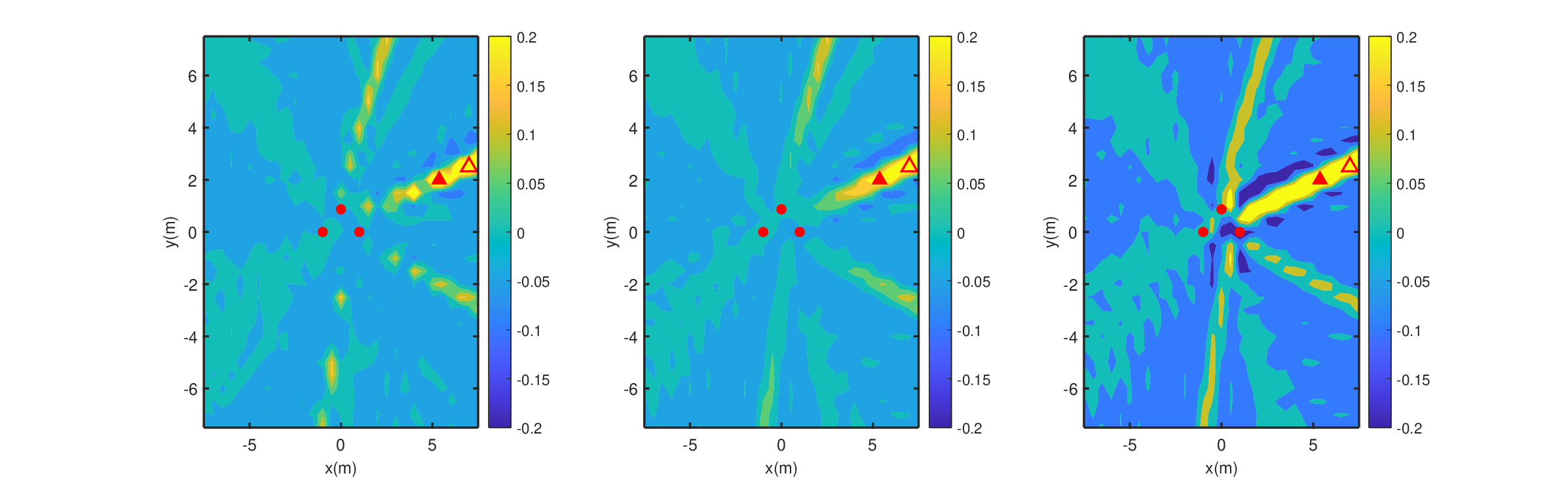}
	\caption{Same as Fig.~\ref{fig:SRP_near}, but with the simulated source further from the microphone array.}
	\label{fig:SRP_far}
\end{figure*}

The qualitative effect of applying the band limitation procedure proposed before is illustrated in Figs. \ref{fig:SRP_near} and \ref{fig:SRP_far}. Both correspond to simulation in fully anechoic conditions of the acoustic propagation of a speech signal taken from the database described in section \ref{subsec:AudioData}. In both cases a triangular microphone array has been supposed, with the sound source placed in the same plane, in a nearby position in Fig.~\ref{fig:SRP_near}, and in a further position in Fig.~\ref{fig:SRP_far}. The left plot in both figures shows the standard SRP-PHAT map, while the middle and right plots show the SRP-PHAT maps calculated with the procedure proposed here, both without (middle) and with (right) the normalization in (\ref{eq:BandlimitedNormalisedGCCPHAT}). One noticeable effect of limiting the band of the GCC is a reduction on the number and relative relevance of the local maxima in the resulting SRP map, which makes it more robust against changes in spatial resolution. For source positions far from the centre of the array, the norm of the TDOA gradient (Fig.~\ref{fig:Gradient}) is low, which results in little band limitation and, consequently, similar results are expected in the estimation of source positions; this is the case illustrated in Fig.~\ref{fig:SRP_far}. However, for source positions near the microphone array, greater differences in the estimated source positions are expected, as the case in Fig.~\ref{fig:SRP_near}.

Calculating SRP maps is computationally expensive, and the fact that calculating band-limited GCCs, as specified in (\ref{eq:BandlimitedGCCPHAT_2}) and (\ref{eq:BandlimitedNormalisedGCCPHAT}), increases such computational cost cannot be overlooked. To present, several strategies have been proposed to speed up SRP calculation, such as decomposing the SRP map in spatial basis functions \cite{DmJA07}, or look-up tables for TDOA values \cite{CYCK09}. These strategies can be extrapolated to the case of using band-limited GCCs by running a spatial analysis before calculating GCCs in order to identify the required bandwidths, computing  and storing GCCs, and using them as look-up tables when building SRP maps. However, it should be stressed that implementation issues are beyond the scope of this research. 

\section{Experiments and results}
\label{sec:Experiments}
The procedure proposed in section \ref{sec:Calculation} to calculate SRP maps has been incorporated into some simulation experiments in order to evaluate its potential impact on the source localization performance of systems based on SRP maps. 
\subsection{Audio data}
\label{subsec:AudioData}
The signals used for the simulation experiments corresponded to several acoustic events included in the \emph{Sound event detection in synthetic audio} task of the DCASE 2016 challenge \cite{DCASE16}. Its associated dataset includes audio files corresponding to 11 types of sound events. According to the spectral analysis reported in \cite{GFCD16}, these types of sound events can be grouped into several categories taking into account their spectra. Specifically, the shape of the spectra can be classified in the following four categories:
\begin{itemize}
	\item \emph{Noisy (non-harmonic) low-pass spectra}, which includes the cases of door slams, opening or closing drawers, typing, door knocking, and page turning.
	\item \emph{Low-pass spectra with resonances due to the human vocal tract}, as in the case of clearing one's throat, coughing, laughing, and speaking.
	\item \emph{Noisy flat spectra}, which is the case for key dropping events.
	\item \emph{Harmonic spectra with flat envelope}, as in phone ringing.
\end{itemize}

In order to cover all the four classes of spectral shape, one event type from each class was selected for running the simulations, namely door slams, speaking, key dropping, and phone ringing. All the 20 recordings corresponding to each type included in the development dataset of the challenge were used, which resulted in a total of 80 recordings. In all cases, the sampling rate was equal to 44.1~kHz and the sound was sampled with a resolution of 16~bits. The duration of the recordings ranged from 0.13~s to 3.34~s.

\subsection{Experiments}
The aforementioned sound events were simulated to happen in a $8\ \mathrm{m}\times 10\ \mathrm{m} \times 4\ \mathrm{m}$ room. Specifically, 1000 source positions inside the room were randomly selected with uniform probability distribution. For each source position one audio recording corresponding to each event type was randomly selected, thus resulting in a total of 4000 simulated sound events. The sound propagation between the source positions and the microphones was simulated by delaying each signal according to the corresponding propagation distance. The sound speed was assumed equal to 343~m/s.

Simulations were carried out for two different microphone arrays. Both of them were formed by 4 microphones placed in the corners of a regular tetrahedron whose central point was located at the centre of the room. The length of the tetrahedron edges were 0.5~m in one case (small array) and 3~m in the other (large array).

The position of the sound source in each case was estimated from the simulated microphone signals according to the algorithm in \ref{sec:Calculation}. The resolution chosen for generating the SRP maps was 0.5~m for all experiments except for one performed with 1~m resolution for the sake of assessing the effect of increasing resolution. The signal bandwidth was assumed to be between 100~Hz and 6000~Hz. According to the analysis in \cite{GFCD16}, the signal-to-noise ratio beyond 6000~Hz was poor for low-pass signals. The localization error was calculated as the absolute value of the difference between the estimated source positions and the actual simulated positions.

Two different acoustic conditions were simulated: anechoic conditions and reverberant conditions with reverberation time equal to 0.6 s, corresponding to a realistic low-reverberant environment \cite{Cowa07}. Simulations of the reverberant room were performed using the image method proposed by Allen and Berkley in \cite{AlBe79}, as implemented in Matlab\textregistered~by Habets \cite{Habe06}.

\subsection{Results}
\begin{figure}[t]
	\centering
	\subcaptionbox{Small array.\label{fig:HIST_errorSmall}}{\includegraphics[width=\linewidth]{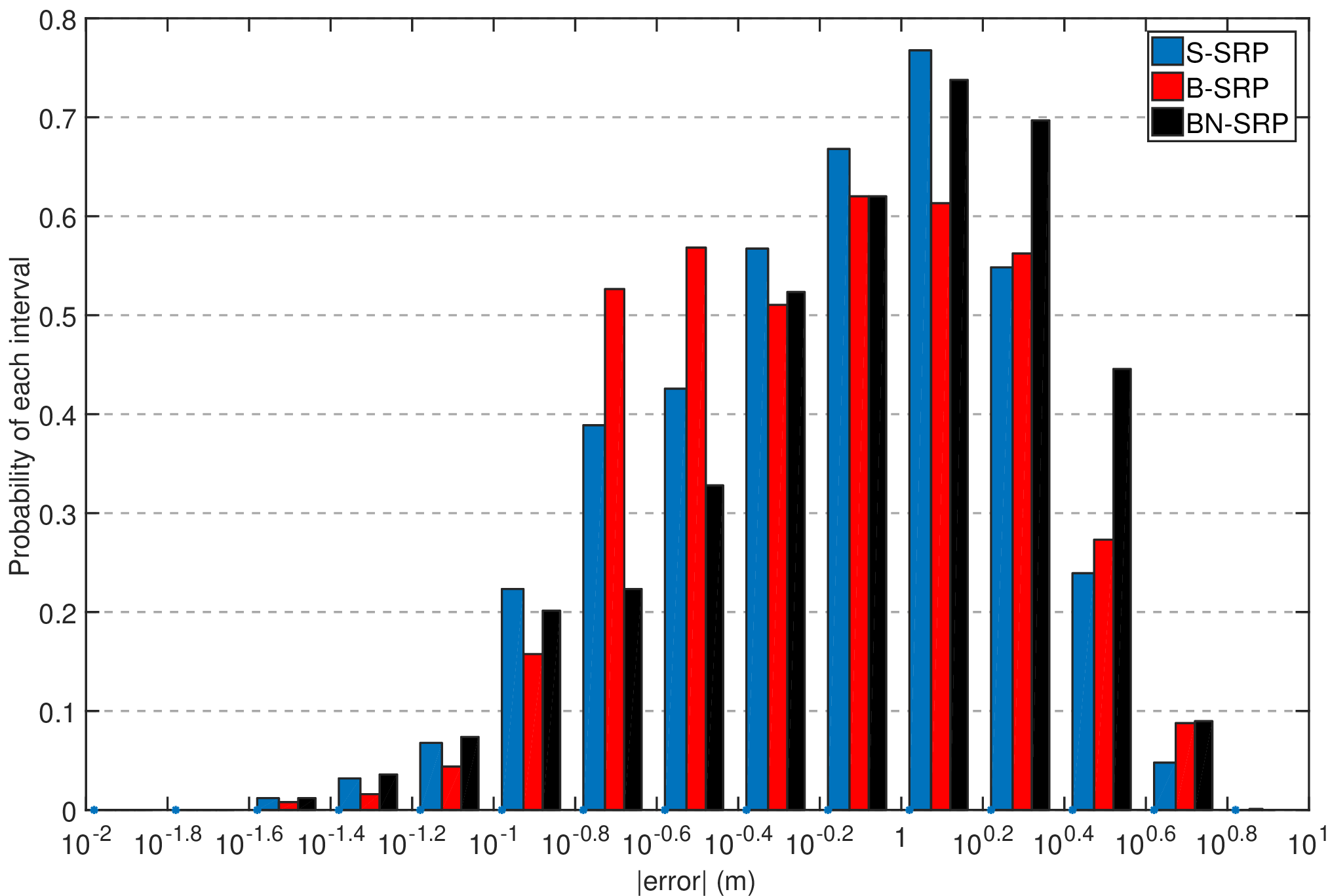}}
	\subcaptionbox{Large array.\label{fig:HIST_errorLarge}}{\includegraphics[width=\linewidth]{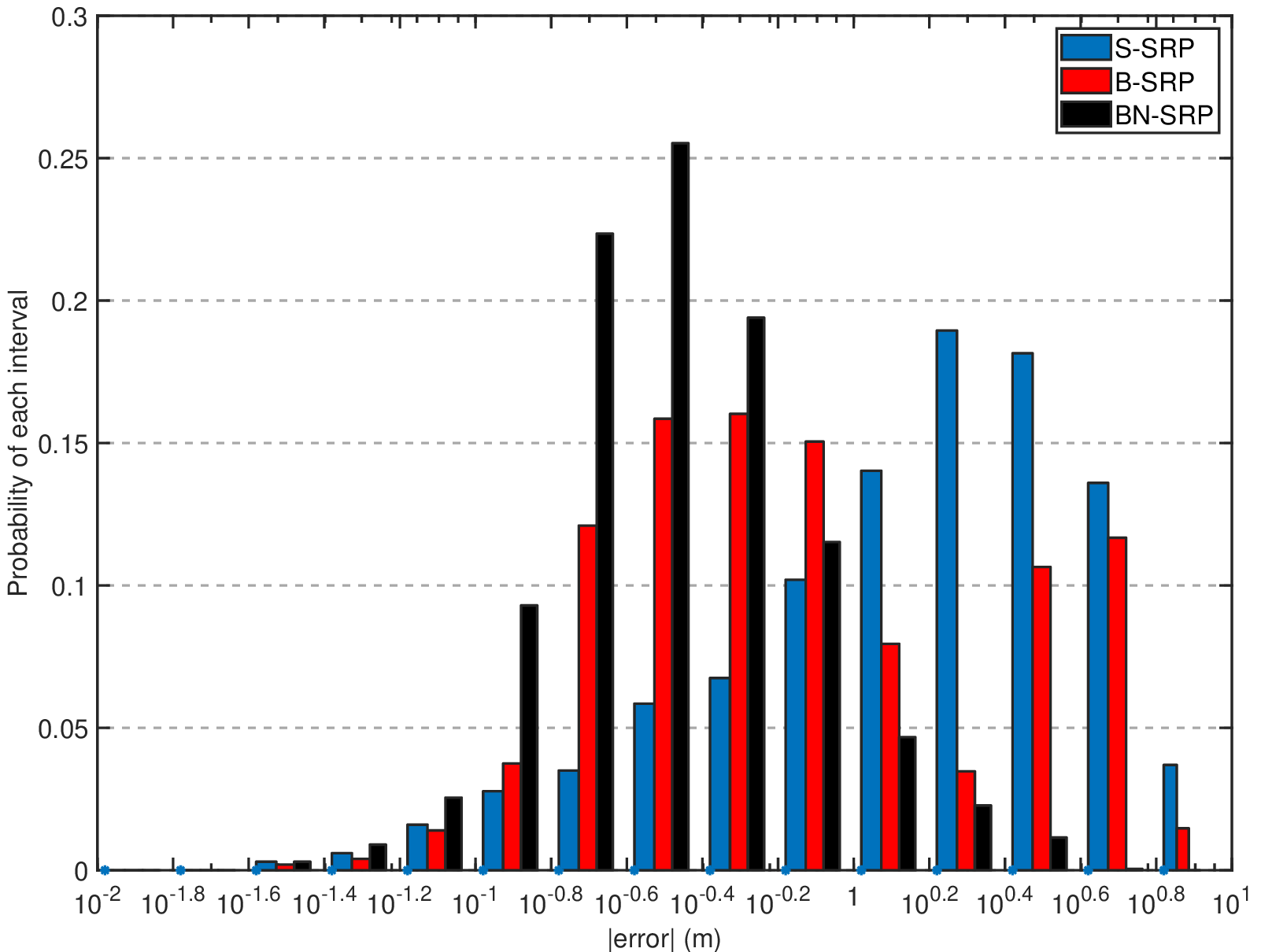}}
	\caption{Histograms of localization errors for the small and large  arrays. Results are given as the probability of each interval in the $x$ axis for SRP maps calculated using the standard GCC (\ref{eq:BandlimitedGCCPHAT}) (S-SRP), the band-limited GCC (\ref{eq:BandlimitedGCCPHAT_2}) (B-SRP), and the normalized band-limited GCC (\ref{eq:BandlimitedNormalisedGCCPHAT}) (BN-SRP).}
	\label{fig:HIST_error}
\end{figure}

Fig.~\ref{fig:HIST_error} shows the histograms representing the distributions of localization errors for the anechoic scenario mentioned before, and for SRP maps calculated using the standard GCC (\ref{eq:BandlimitedGCCPHAT}) (S-SRP), the band limited GCC (\ref{eq:BandlimitedGCCPHAT_2}) (B-SRP), and the normalized band-limited GCC (\ref{eq:BandlimitedNormalisedGCCPHAT}) (BN-SRP). The plot in Fig.~\ref{fig:HIST_errorSmall} shows the histograms for the small array, while the plot in Fig.~\ref{fig:HIST_errorLarge} corresponds to the large array. At first sight, the use of the band-limited GCC does not produce results significantly different to those of the standard GCC. Furthermore, the normalization proposed in (\ref{eq:BandlimitedNormalisedGCCPHAT}) produces a moderate worsening of the localization performance. But in the case of the large array (Fig.~\ref{fig:HIST_errorLarge}) the band limitation in the GCC produces a relevant reduction in localization error and the magnitude of this reduction is higher when the normalization in (\ref{eq:BandlimitedNormalisedGCCPHAT}) is applied. 

Table \ref{tab:AnechoicErrors} provides a quantitative description of the distributions of localization errors. Specifically, the average error for each case, and the mean deviation of errors are given. It is apparent from Fig. \ref{fig:HIST_error} that distributions cannot be assumed to be Gaussian (e.g. the distributions of errors for B-SRP is bimodal). For this reason, nonparametric tests were chosen to evaluate the statistical significance of differences among the means and the dispersions (mean deviations) of distributions. Namely, the Wilcoxon test was used to evaluate differences in the mean values, and a permutation test of deviances to evaluate differences in the dispersions \cite{Higg04}. For a given distribution of $N$ estimation errors $e_n$, the mean or average error is defined as:
\begin{equation}
	\mu_e=\frac{1}{N}\sum_{n=1}^N e_n
\end{equation}
and the mean deviation is:
\begin{equation}
	\delta_e=\frac{1}{N}\sum_{n=1}^N \left|e_n -  \mu_e \right| .
\end{equation}

\begin{table}
	\caption{Average value and mean deviation of the distributions of localization errors for the anechoic scenario.}
	\label{tab:AnechoicErrors}
	\setlength{\tabcolsep}{3pt}
	\begin{tabularx}{\linewidth}{m{1cm} m{1cm} m{1cm} m{1cm} m{1cm} m{1cm} m{1cm}}
		\toprule[1pt]
		 & \multicolumn{3}{c}{\textbf{Mean error (m)}}& \multicolumn{3}{c}{\textbf{Mean deviation (m)}} \\
		 \cmidrule[0.75pt](r{0.3em}){2-4}
		 \cmidrule[0.75pt](l{0.3em}){5-7}
		 & S-SRP & B-SRP & BN-SRP & S-SRP & B-SRP & BN-SRP \\
		 \midrule
		 Small array & 1.024$^*$ & 1.062$^*$  & 1.268 & 0.6845 & 0.7653 & 0.8308 \\
		 \midrule
		 Large array & 2.2405 & 1.478  & 0.4693 & 1.485 & 1.383 & 0.2929 \\
		 \bottomrule[1pt]
		 \addlinespace[1pt]
		 \multicolumn{7}{l}{The difference between quantities marked with $^*$ is not statistically}\\
		 \multicolumn{7}{l}{significant ($p>0.01$).}
	\end{tabularx}
\end{table}
Results in Tab. \ref{tab:AnechoicErrors} confirm the observations that S-SRP and B-SRP perform similarly for the small array (non significant differences in the mean errors and similar values for dispersions) but not for the large array, B-SRP performs better in that case, and BN-SRP provides a significant performance improvement for the large array, while it performs poorly for the small array.

A deeper insight into the previous results can be obtained if they are segmented by the distance of the simulated sources to the centre of the array. This can be done using the relative measure $r / r_\mathrm{m}$. Table \ref{tab:AnechoicErrorsPerDistance} shows the average errors and their mean deviances discriminated for three intervals: $r / r_\mathrm{m} \le 5$; $5 < r / r_\mathrm{m} \le 10$; and $10 < r / r_\mathrm{m}$. These results indicate that the performance of all three algorithms is not as dependent from the array size as from the relative distance between the source and the array centre. Note that all 1000 simulated source positions comply with the condition $r / r_\mathrm{m} \le 5$ in the case of the large array, while the distribution for the small array is:
\begin{itemize}
	\item 40 points in the $r / r_\mathrm{m} \le 5$ interval,
	\item 274 points in the $5 < r / r_\mathrm{m} \le 10$ interval,
	\item and 686 points in the $10 < r / r_\mathrm{m}$ interval.  
\end{itemize}

\begin{table}
	\caption{Average value and mean deviation of the distributions of localization errors for the anechoic scenario discriminated for three diferent distance intervals.}
	\label{tab:AnechoicErrorsPerDistance}
	\setlength{\tabcolsep}{2.5pt}
	\begin{tabularx}{\linewidth}{ m{2cm} m{0.775cm} m{0.8cm} m{1cm} m{0.775cm} m{0.8cm} m{1cm}}
		\toprule[1pt]
	    & \multicolumn{3}{c}{\textbf{Mean error (m)}}& \multicolumn{3}{c}{\textbf{Mean deviation (m)}} \\
		\cmidrule[0.75pt](r{0.3em}){2-4}
		\cmidrule[0.75pt](l{0.3em}){5-7}
		 & S-SRP & B-SRP & BN-SRP & S-SRP & B-SRP & BN-SRP \\
		\midrule
		Small array \par $\left(\frac{r}{r_\mathrm{m}} \le 5\right)$& 2.149 & 3.210 & 0.8203 & 1.151$^*$ & 1.196$^*$ & 0.6731 \\
		\midrule
		Large array \par $\left(\frac{r}{r_\mathrm{m}} \le 5\right)$& 2.241 & 1.478 & 0.4693 & 1.485 & 1.383 & 0.2929 \\
		\midrule
	 	Small array \par $\left(5 < \frac{r}{r_\mathrm{m}} \le 10\right)$& 1.313 & 1.637 & 1.049 & 0.8179 & 0.9324 & 0.5759 \\
		\midrule
		Small array \par $\left(10 < \frac{r}{r_\mathrm{m}}\right)$& 0.8428 & 0.7066 & 1.3811 & 0.5433 & 0.4536 & 0.9285 \\
		\bottomrule[1pt]
        \addlinespace[1pt]
		\multicolumn{7}{l}{The difference between quantities marked with $^*$ is not statistically}\\
		\multicolumn{7}{l}{ significant ($p>0.01$).}
	\end{tabularx}
\end{table}

It can be observed that the localization error using the standard GCC diminishes as the distance between the centre of the array and the source position is increased. The performance of the estimator based on the band-limited GCC is also increased for longer distances, but this estimator differs from the previous one mainly in two aspects: the average reduction in the localization error is greater as distance is increased, and the dispersion of the localization errors is also the lowest for the longest evaluated distances. The SRP map based on the band-limited and normalized GCC provides estimations that behave oppositely with respect to the other two cases. This method provides better estimations for short distances between array and sound source, and its performance is negatively affected by increases in these distances.

The results of running the same experiments but with reverberation time equal to 0.6~s (Tab. \ref{tab:ReverbErrorsPerDistance}) indicate that, in general terms, the presence of reverberation tends to negatively affect localization results. In fact, all mean errors in Tab. \ref{tab:ReverbErrorsPerDistance} are higher than the corresponding values in Tab. \ref{tab:AnechoicErrorsPerDistance}, except for those relative to S-SRP and B-SRP being applied to the few points with $r / r_\mathrm{m} \le 5$ in the small array case. Such increase in average error happens more prominently for BN-SRP. Another relevant aspect of these results is that the growth in average error is less relevant for the most distant sources ($10 < r/r_\mathrm{m}$), and that B-SRP still provides the best performance for this case in the reverberant scenario.

\begin{table}
	\caption{Average value and mean deviation of the distributions of localization errors for the reverberant scenario discriminated for three diferent distance intervals.}
	\label{tab:ReverbErrorsPerDistance}
	\setlength{\tabcolsep}{2.5pt}
	\begin{tabularx}{\linewidth}{ m{2cm} m{0.775cm} m{0.8cm} m{1cm} m{0.775cm} m{0.8cm} m{1cm}}
		\toprule[1pt]
		& \multicolumn{3}{c}{\textbf{Mean error (m)}}& \multicolumn{3}{c}{\textbf{Mean deviation (m)}} \\
		\cmidrule[0.75pt](r{0.3em}){2-4}
		\cmidrule[0.75pt](l{0.3em}){5-7}
		& S-SRP & B-SRP & BN-SRP & S-SRP & B-SRP & BN-SRP \\
		\midrule
		Small array \par $\left(\frac{r}{r_\mathrm{m}} \le 5\right)$& 2.222 & 3.193 & 0.9436 & 1.196$^*$ & 1.204$^*$ & 0.6189 \\
		\midrule
		Large array \par $\left(\frac{r}{r_\mathrm{m}} \le 5\right)$& 3.009 & 3.306$^\diamond$ & 3.111$^\diamond$ & 1.770 & 2.112 & 1.575 \\
		\midrule
		Small array \par $\left(5 < \frac{r}{r_\mathrm{m}} \le 10\right)$& 1.263 & 1.638 & 2.028 & 0.7881 & 0.9091 & 0.4071 \\
		\midrule
		Small array \par $\left(10 < \frac{r}{r_\mathrm{m}}\right)$& 0.9822 & 0.8276 & 3.976 & 0.6682 & 0.5754 & 0.6686 \\
		\bottomrule[1pt]
		\addlinespace[1pt]
		\multicolumn{7}{l}{The differences between quantities marked with $^*$, and $^\diamond$ are not }\\
		\multicolumn{7}{l}{statistically significant ($p>0.01$).}
	\end{tabularx}
\end{table}

The effect of changing map resolution on the relative performance of all three options for calculating the GCC was assessed by running one experiment with coarser resolution (1~m) in the reverberant scenario. Considering the poor localization performance for short distances in this scenario (Tab.~\ref{tab:ReverbErrorsPerDistance}), only the small array was simulated in this case. Results summarized in Tab.~\ref{tab:CoarseErrorsPerDistance} show that the advantages provided by band-limiting the GCC, either with or without normalization, are more noteworhty in this case. In other words, performance of the SRP based on the standard GCC seems to be more sensitive to increases in map resolution than that of SRP based on the band-limited GCC. 

\begin{table}
	\caption{Average value and mean deviation of the distributions of localization errors for the reverberant scenario and the small array with coarser map resolution (1 m).}
	\label{tab:CoarseErrorsPerDistance}
	\setlength{\tabcolsep}{2.5pt}
	\begin{tabularx}{\linewidth}{ m{2cm} m{0.775cm} m{0.8cm} m{1cm} m{0.775cm} m{0.8cm} m{1cm}}
		\toprule[1pt]
		& \multicolumn{3}{c}{\textbf{Mean error (m)}}& \multicolumn{3}{c}{\textbf{Mean deviation (m)}} \\
		\cmidrule[0.75pt](r{0.3em}){2-4}
		\cmidrule[0.75pt](l{0.3em}){5-7}
		& S-SRP & B-SRP & BN-SRP & S-SRP & B-SRP & BN-SRP \\
		\midrule
		Small array \par $\left(\frac{r}{r_\mathrm{m}} \le 5\right)$& 2.565 & 2.957 & 0.9977 & 1.006$^*$ & 1.080$^*$ & 0.5866 \\
		\midrule
		Small array \par $\left(5 < \frac{r}{r_\mathrm{m}} \le 10\right)$& 2.056$^\diamond$ & 1.840$^\dagger$ & 1.7016$^{\diamond,\dagger}$ & 1.110$^\ddagger$ & 1.020$^\ddagger$ & 0.4139 \\
		\midrule
		Small array \par $\left(10 < \frac{r}{r_\mathrm{m}}\right)$& 1.665 & 0.9344 & 3.651 & 1.150 & 0.5468 & 0.6794 \\
		\bottomrule[1pt]
		\addlinespace[1pt]
		\multicolumn{7}{l}{The differences between quantities marked with $^*$, $^\diamond$, $^\dagger$, and $^\ddagger$, are not }\\
		\multicolumn{7}{l}{statistically significant ($p>0.01$).}
	\end{tabularx}
\end{table}

\section{Discussion and conclusions}
\label{sec:Discussion}
The analysis presented in sections \ref{sec:Geometrical analysis} and \ref{sec:Implications} was aimed at calculating SRP maps avoiding the potential aliasing effects that may happen when sampling the GCC function regardless the relation between SRP map resolution and GCC bandwidth. This analysis led to the sufficient condition (\ref{eq:SamplingCondition2}) that allows avoiding such aliasing. However, the inequality in (\ref{eq:SamplingTime}) implies that fulfilling this condition is not necessary to avoid aliasing or, in other words, that by applying this condition one can limit the bandwidth of the GCC more than what is strictly necessary. As a consequence, the localization errors produced by B-SRP may be sometimes larger than those of S-SRP, as can be noticed in the histogram in Fig.~\ref{fig:HIST_errorSmall}. According to the numerical results summarized in Tabs. \ref{tab:AnechoicErrorsPerDistance} and \ref{tab:ReverbErrorsPerDistance}, this worsening of localization performance happens especially for the shortest distances between microphone array and sound source position ($r / r_\mathrm{m} \le 5$). 

One possible explanation for the afore-mentioned worsening of localization performance was hypothesized to be the reduction in the height of the main peak of the GCC that is intrinsically linked to bandwidth limitation (see Fig.~\ref{fig:GCC}). As illustrated in Figs. \ref{fig:GradientMap} and \ref{fig:Gradient}, points near the microphones are associated to the highest TDOA gradients and, consequently, the calculation of their corresponding SRP values is affected the most by the bandwidth limitation of the GCC. This implies a reduction of the height of the main GCC peak and a corresponding reduction of the SRP value. The GCC normalization in (\ref{eq:BandlimitedNormalisedGCCPHAT}) was proposed to compensate this effect at the cost of increasing the amplitude of some secondary GCC peaks (Fig.~\ref{fig:GCC}). The inclusion of this normalization factor in the calculation of SRP maps has shown to have a very positive impact on localization performance for sound source positions near or even inside the volume occuppied by the microphone array (see Fig.~\ref{fig:HIST_errorSmall} and Tab.~\ref{tab:AnechoicErrorsPerDistance}). However, the effect of such normalization on the secondary peaks in the GCC (Fig.~\ref{fig:GCC}) is likely to be a key factor in worsening the performance of this approach for reverberant environments (Tab.~\ref{tab:ReverbErrorsPerDistance}).

For longer source-to-array distances ($10 < r / r_\mathrm{m}$), applying bandwidth limitation to the GCC according to (\ref{eq:SamplingCondition2}) has shown to consistently provide performance improvements over the standard approach for calculating SRP without limiting the bandwidth of the GCC (Tabs.~\ref{tab:AnechoicErrorsPerDistance} and \ref{tab:ReverbErrorsPerDistance}). These improvements involve reductions in both average error and error dispersion. The reason that justifies the improved performance of B-SRP can be explained by looking back to Fig.~\ref{fig:SRP_near}. The left map therein shows a typical localization error of S-SRP maps. This error is mainly caused by two factors: the calculation of the SRP map misses a relevant GCC peak near the actual source position due to this peak being narrower than the corresponding map resolution, and some secondary GCC peaks are added up in a position nearer the centre of the array, hence producing a peak in the SRP map higher than it should be. As illustrated by Fig.~\ref{fig:GCC}, limiting the bandwidth of the GCC has the double effect of widening the main GCC peak and eliminating some secondary peaks. This has the consequence of avoiding errors in which the distance between the sound source and the microphone array is underestimated.

\begin{figure*}[t]
	\subcaptionbox{Small array.\label{subfig:Distance2distanceSmall}}{\includegraphics[width=\linewidth]{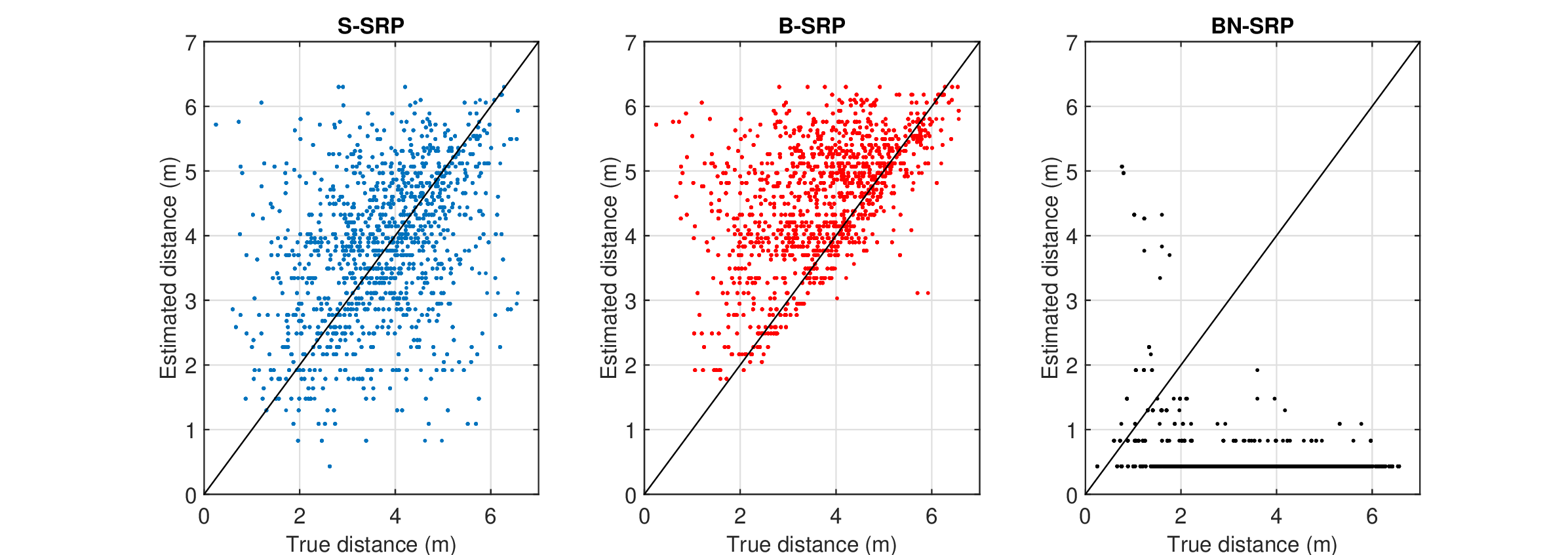}}
	\subcaptionbox{Large array.\label{subfig:Distance2distanceLarge}}{\includegraphics[width=\linewidth]{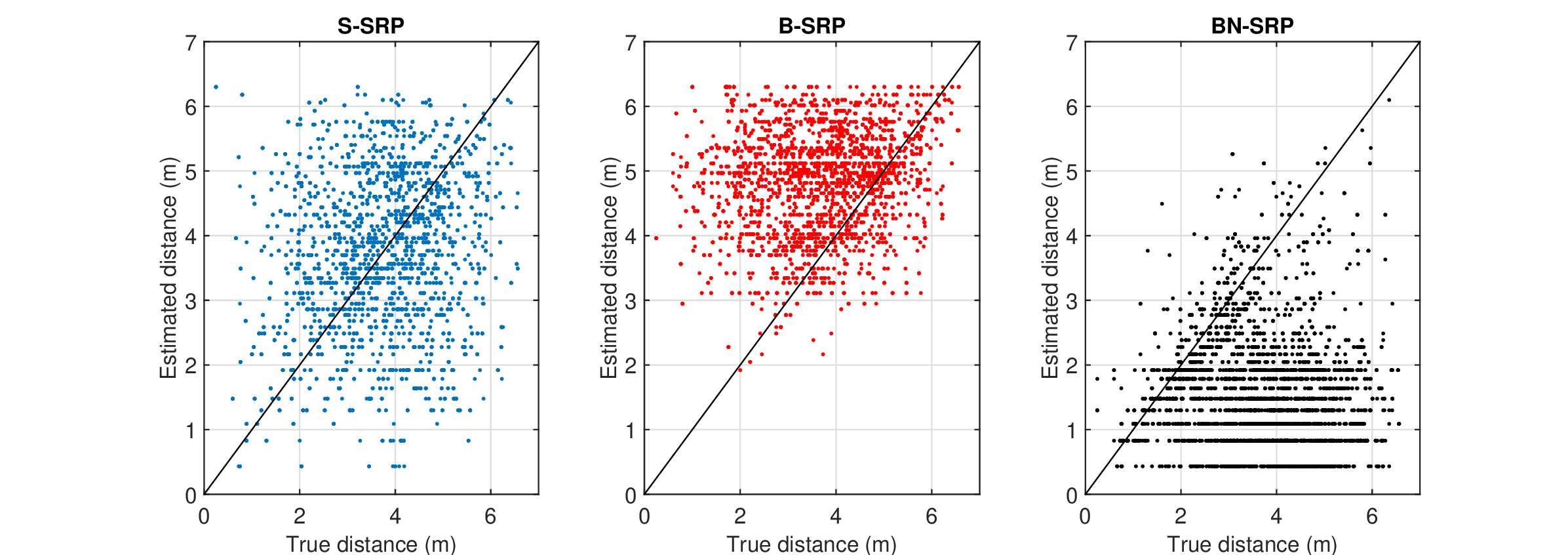}}
	\caption{Scatter plots showing the relation of the actual distance between the sound source and the centre of the microphone array to the distance between the estimated source position and the centre of the array. Plots correspond, left to right, to S-SRP (SRP based on the standard GCC), B-SRP (SRP based on the band-limited GCC), and BN-SRP (SRP based on the band-limited normalized GCC).}
	\label{fig:Distance2distance}
\end{figure*}

The left and central scatter plots in Fig.~\ref{fig:Distance2distance} represent the relation between the real source-to-array distance and the estimated source-to-array distance for both S-SRP and B-SRP in the case of the reverberant scenario. These plots show that the previously mentioned consequence of limiting the bandwidth of the GCC does not only happen in specific points; instead, it is a general rule for the results of our experiments that limiting the bandwidth of the GCC according to (\ref{eq:SamplingCondition2}) reduces the probability of underestimating source-to-array distances. This explains why the B-SRP performs better for large distances (Tabs.~\ref{tab:AnechoicErrorsPerDistance} and \ref{tab:ReverbErrorsPerDistance}). When source-to-array distances are significantly larger than array size, underestimating this distance is an issue, and B-SRP performs the best for the largest simulated distances. 

A reasoning analogous to the previous one leads to the conclusion that the GCC normalization in (\ref{eq:BandlimitedNormalisedGCCPHAT}) has the effect of increasing the values of the SRP map in positions near the centre of the array. Thus, it reduces the chance of overestimating the source-to-array distance. This effect is confirmed by the right plot in Fig.~\ref{fig:Distance2distance}. However, the presence of reverberation has a very negative impact on localization performance when source-to-array distances are in the range of, or even shorter than the array size ($r / r_\mathrm{m} \le 5$), which corresponds to the case where the microphones are more distributed in the room. Thus, the negative impact of reverberation masks the potential benefits of using BN-SRP in reverberant scenarios. Yet, note that even in this case BN-SRP yields significantly lower error dispersion for that range of distances than both S-SRP and B-SRP (Tab.~\ref{tab:ReverbErrorsPerDistance}) for similar average errors.

The high computational cost of calculating SRP with fine spatial resolutions has led several researchers to propose iterative approaches to sound source localization, consisting in a step-by-step cecrease in map resolution accompanied by a corresponding reduction in map extent, as mentioned in the introduction. The analysis presented in this paper has made no assumption about specific intervals for map resolution, so it is applicable at any scale in those iterative or hierarchical approaches. To illustrate this, an additional experiment was run with map resolution equal to 1~m instead of 0.5~m. The corresponding results, summarized in Tab. ~\ref{tab:CoarseErrorsPerDistance}, confirmed all the previously stated conclusions. Furthermore, the increased map resolution implies the requirement of a narrower spectrum according to (\ref{eq:SamplingCondition2}) or, from another point of view, coarser resolutions in SRP maps lead to more relevant aliasing effects if the bandwidth of the GCC is not limited. Such increased aliasing leads to a noticeable worsening of localization results for S-SRP (compare results in Tab.~\ref{tab:ReverbErrorsPerDistance}  to those in Tab.~\ref{tab:CoarseErrorsPerDistance}). However, the impact of the coarser resolution in the performance of B-SRP and BN-SRP is much lower, to the extent that B-SRP provides significantly better results than S-SRP even for intermediate distances ($5 < r / r_\mathrm{m} \le 10$), which was not the case when the resolution was 0.5~m.

In conclusion, equations (\ref{eq:SamplingTime}) and (\ref{eq:SamplingCondition1}) show that there is a relation between the bandwidth of acoustic signals and the resolution of SRP-PHAT maps calculated for localizing their corresponding source. This relation implies the sufficient condition for an aliasing-free calculation of the SRP map specified by (\ref{eq:SamplingCondition2}). Such calculation can be done according to the algorithm described in section \ref{sec:Calculation} and limiting the bandwidth of the GCC as indicated in (\ref{eq:BandlimitedGCCPHAT_2}). While the fact that integrating (i.e. low-pass filtering) the GCC leads to increased robustness in localization performance of SRP-PHAT maps was already known \cite{CoML11}, the analysis presented before provides a theoretical justification for such improvement and an explicit rule that relates GCC bandwidth to the spatial resolution of SRP-PHAT maps.

The reported experiments show that this approach leads to improved source localization results for source positions far from the microphone array, since the probability of underestimating the source-to-array distance is reduced. It has also been tested that the proposed approach is robust against reverberation, since it provides similar advantages in both anechoic and reverberant scenarios. Last, it should be stressed that the use of condition  (\ref{eq:SamplingCondition2}) to avoid aliasing effects in the calculation of SRP maps is fully compatible with hierarchical localization algorithms in which map resolution is iteratively changed. Moreover, it should significantly contribute to obtain more robust results at the coarsest resolution levels.

\bibliographystyle{IEEETRAN}
\bibliography{IEEEABRV,TDOA}

\begin{IEEEbiography}[{\includegraphics[width=1in,height=1.25in,clip,keepaspectratio]{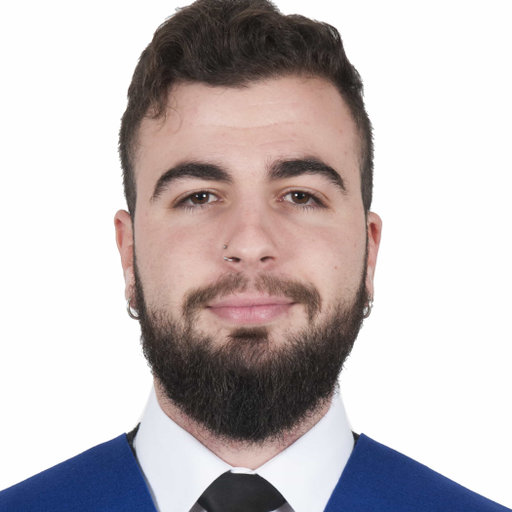}}]{Guillermo Garc\'{\i}a-Barrios} was born in Madrid (Spain) in 1994. He received the B.S. degree in sound and image engineering from the Universidad Polit\'ecnica de Madrid (UPM) in 2017, and the M.S. degree in systems and services engineering for the information society from the same university in 2018. He is currently doing a Ph.D. in acoustic signal processing at the UPM.

His research interest includes sound source localization algorithms and systems, acoustic simulation, machine learning for automatic sound signal recognition, and wireless acoustic sensor networks.
\end{IEEEbiography}

\begin{IEEEbiography}[{\includegraphics[width=1in,height=1.25in,clip,keepaspectratio]{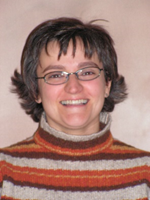}}]{Juana M. Guti\'errez-Arriola} (M'12) was born in Santander (Spain) in 1971. She received the B.S. and M.S. degrees in telecommunications engineering from the Universidad de Cantabria in 1993 and 1994 respectively. She did her Ph.D. at the Universidad Polit\'ecnica de Madrid (UPM), completing it in 2008. 
	
Currently, she is a Senior Lecturer of the Telematics and Electronics Engineering department of the UPM. She belongs to the research group on Acoustics and Multimedia Applications (GAMMA) of the UPM, where she carries out research on biomedical image processing, speech analysis and sound source localization. She has authored over 25 scientific papers in those fields.

Dr. Guti\'errez-Arriola has been Secretary of the Escuela Universitaria de Ingenier\'{\i}a T\'ecnica de Telecomunicaci\'o in the UPM from 2004 until 2008, Head of the Department of Circuits and Systems Engineering (UPM) from 2011 to 2014, and she currently is Secretary of the Research Center on Software Technologies and Multimedia Systems for Sustainability, also in the UPM.
\end{IEEEbiography}

\begin{IEEEbiography}[{\includegraphics[width=1in,height=1.25in,clip,keepaspectratio]{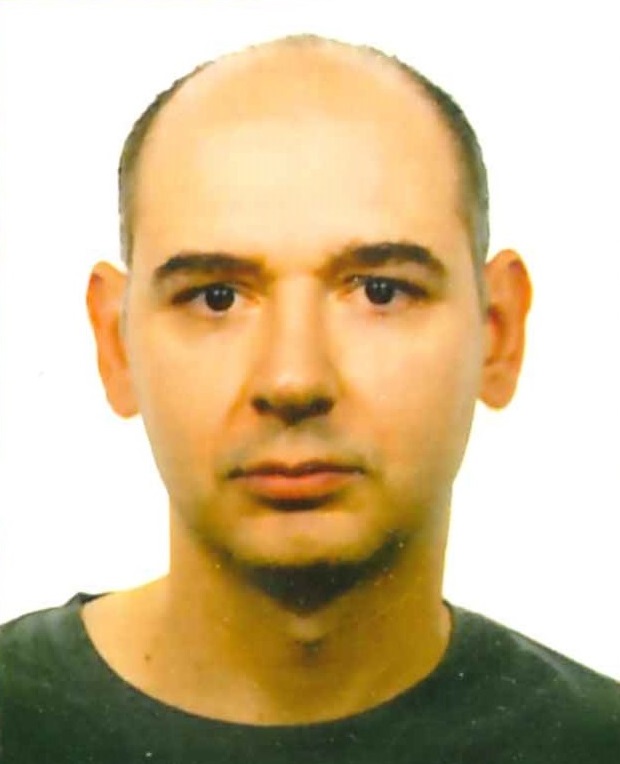}}]{Nicol\'as S\'aenz-Lech\'on} was born in Barcelona (Spain) in 1972. He received his B.S. (1996), M.S. (2001), and Ph.D. (2010) degrees in telecommunications engineering from the Universidad  Polit\'ecnica de Madrid (UPM) in Spain.
	
Since 2019, he is Lecturer at the department of Audiovisual Engnieering and Communications at the UPM. His main research interests are biomedical image processing, and speech analysis for clinical applications. He is the author of over 20 journal papers and more than 50 conference papers in those fields. 
\end{IEEEbiography}

\begin{IEEEbiography}[{\includegraphics[width=1in,height=1.25in,clip,keepaspectratio]{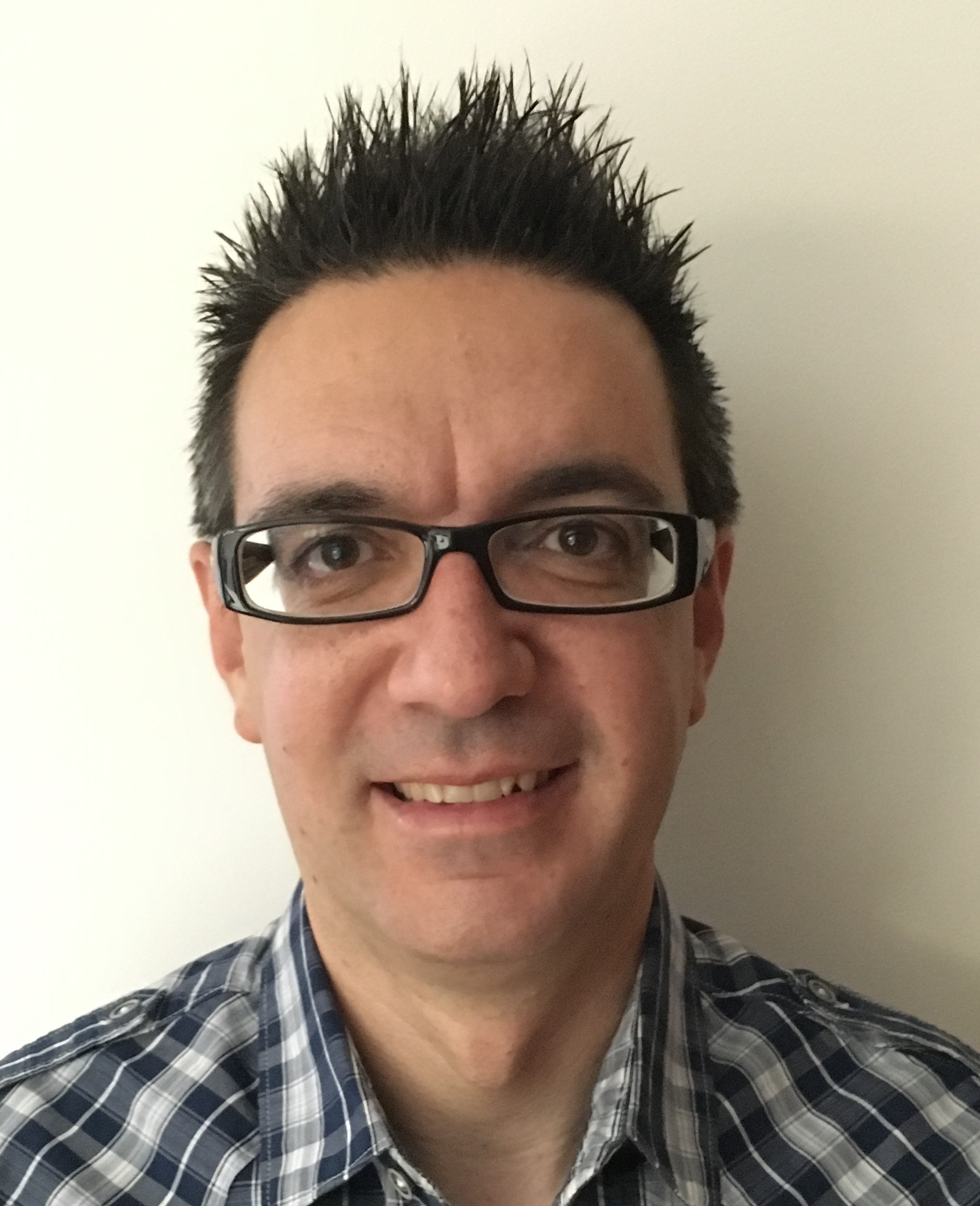}}]{V\'{\i}ctor Jos\'e Osma-Ruiz} was born in Cuenca (Spain) in 1973. He received his  B.S. (1995), M.S. (2001) and Ph.D. (2010) degrees in telecommunications engineering from the Universidad  Polit\'ecnica de Madrid (UPM) in Spain.

He gives lectures at the Escuela T\'ecnica Superior de Ingenier\'{\i}a y Sistemas de Telecomunicaci\'on in the UPM since 1999, where he currently holds a Senior Lecturer position at the Telematics and Electronics Engineering department. His research interest includes biomedical image processing, speech analysis for clinical applications, gamification, and virtual reality. He has published 20 journal papers and another 60 scientific papers in conferences and congresses in those fields.
\end{IEEEbiography}

\begin{IEEEbiography}[{\includegraphics[width=1in,height=1.25in,clip,keepaspectratio]{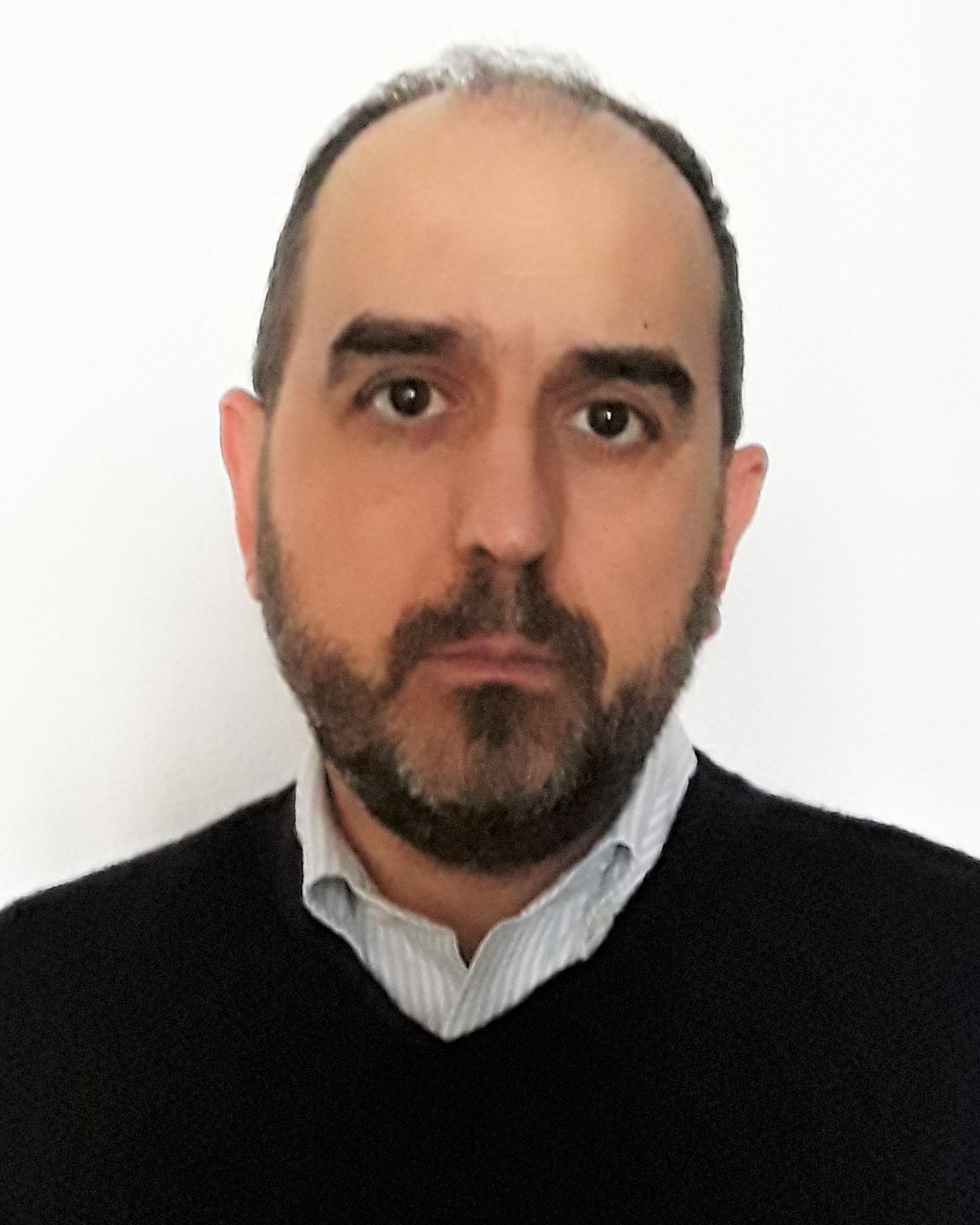}}]{Rub\'en Fraile} was born in Baracaldo (Spain) in 1972. He received his degree in telecommunications engineering (M.S. level) from the Universidad Polit\'ecnica de Valencia (UPV) in 1995. He later did a Ph.D. in mobile communications, receiving his degree from the UPV in 2000. 
	
Before completing his doctorate, he worked as a radio-planning engineer for Retevisi\'on M\'ovil (Spanish mobile network operator), and he later taught several subjects in a secondary school from 2000 to 2007. Simultaneously, he stayed as a postdoctoral researcher in the UPV since 2001 until 2007. During this period, he made research in the simulation of wireless communication networks. He got a position as Senior Lecturer in the Universidad Polit\'ecnica de Madrid (UPM) in 2007, after which he re-oriented his research interests towards the field of discrete-time signal processing applications. He belongs to the research group on Acoustics and Multimedia Applications (GAMMA) of the UPM.

Dr Fraile was the coordinator of the scientific programme in the international workshop on Advanced Voice Function Assessment (AVFA) in 2009, and during his carrer has held two management positions in educational institutions: he coordinated the quality management system (ISO-9000) in a secondary school since 2004 until 2006; and from 2010 to 2013 he was adjunct to the Vice-principal for Academic Staff at the CEU Cardenal Herrera University in Valencia (Spain).
\end{IEEEbiography}

\EOD

\end{document}